\documentclass[journal,12pt,onecolumn]{IEEEtran}
\usepackage{setspace}
\usepackage{amsmath}
\usepackage{comment}
\usepackage{graphicx}
\usepackage[flushleft]{threeparttable} % add footnote in table
\usepackage{subcaption}
\usepackage{tikz}%add lines in subfigures
\usepackage{array}%set column size in table
\usepackage{amssymb} %add therefore in equation
\usepackage{stfloats}
\usepackage{cite}
\usepackage{multirow}
\usepackage[symbol]{footmisc}
\usepackage{booktabs} %tick top and bottom lines in table
\usepackage{tabularx}
\usepackage{caption}
\usepackage{lineno}
\usepackage{float} %force table to be in line at the place
\usepackage{longtable}
\usepackage{url}
\newcommand*{\update}{\textcolor{black}}

\begin{document}

\title{Data-aided Underwater Acoustic Ray Propagation Modeling}

\author{\IEEEauthorblockN{Kexin Li, Mandar Chitre}\\
\IEEEauthorblockA{\update{ARL, Tropical Marine Science Institute and Department of Electrical and Computer Engineering,
National University of Singapore}}}

\maketitle

\IEEEpubid{\begin{minipage}{\textwidth}\ \\[45pt] \centering
\linespread{0.6}\selectfont
\copyright 2023 IEEE. Personal use of this material is permitted. Permission from IEEE must be obtained for all other uses, in any current or future media, including reprinting/republishing this material for advertising or promotional purposes, creating new collective works, for resale or redistribution to servers or lists, or reuse of any copyrighted component of this work in other works.
\end{minipage}}

\begin{abstract}
Acoustic propagation models are widely used in numerous oceanic and underwater applications. Most conventional models are approximate solutions of the acoustic wave equation, and require accurate environmental knowledge to be available beforehand. Environmental parameters may not always be easily or accurately measurable. While data-driven techniques might allow us to model acoustic propagation without the need for extensive prior environmental knowledge, such techniques tend to be data-hungry and often infeasible in oceanic applications where data collection is difficult and expensive. We propose a data-aided \update{physics-based high-frequency} acoustic propagation modeling approach that enables us to train models with only a small amount of data. The proposed framework is not only data-efficient, but also offers flexibility to incorporate varying degrees of environmental knowledge, and generalizes well to permit extrapolation beyond the area where data \update{were} collected. We demonstrate the feasibility and applicability of our method through four numerical case studies, and one controlled experiment. We also benchmark our method's performance against \update{two} classical data-driven techniques \update{-- Gaussian process regression and deep neural network}.
\end{abstract}


\section{Introduction}
\update{Acoustic propagation in typical ocean environments exhibits strong spatial and temporal variability. The dynamic nature of oceans across all spatiotemporal scales poses a difficult problem for underwater acoustics. The ability to effectively model acoustic propagation is vital in many oceanic applications, and therefore necessary although often challenging. Applications related to oceanic acoustic propagation modeling can be broadly categorized into two main classes -- \emph{forward} problems and \emph{inverse} problems. The forward problems seek to estimate acoustic fields at various receiver locations given environmental information~\cite{james2008method, gul2017underwater,llor2012underwater}. On the other hand, inferring unknown environmental parameters from acoustic measurements is the goal of inverse problems~\cite{Chapman2008, chapman2003benchmarking, dosso1993estimation,bonnel2010estimation}. Common inverse problems mainly focus on localization and remote sensing of the ocean environments~\cite{collins1994inverse}. Matched field processing~(MFP) is a generalized beamforming method that is widely used in inverse problems. Conventional MFP employs acoustic propagation modeling to search for environmental parameters that generate field replicas matching acoustic measurements~\cite{tolstoy2000applications, baggeroer1988matched, baggeroer1993overview}}.

\update{Physics-based acoustic propagation models estimate acoustic field by leveraging our physical understanding of acoustic propagation.} Most physics-based models are derived from the acoustic wave equation~\cite{Jensen2011propagation}. Closed-form \update{solutions of} the wave equation are typically analytically intractable in ocean environments~\cite{10.3389/fmars.2021.751327}. There are various widely used approximate solutions to the wave equation, most of which can be seen as variants of the following four groups: ray methods~\cite{Jensen2011}, normal modes~\cite{Jensen2011nm}, parabolic equations~\cite{Jensen2011pb}, and wavenumber integration~\cite{Jensen2011wni}.

There are two main limitations of the \update{physics-based} models -- high computational complexity \update{(especially in complex three-dimensional~(3D) environments)}, and the need for accurate environmental knowledge. A common approach today is to approximate 3D propagation effects by applying \update{two-dimensional}~(2D) models to $N$ azimuths. This idea is often referred \update{to} as 2.5D models or $N \times$2D models in literature~\cite{williamson1995critical, reeder2010experimental}. \update{While it is computationally more efficient than full 3D modeling}, 2.5D models can only be applied in environments where the out-of-plane arrival energy is insignificant. \update{Since the multipath structure of an ocean environment leads to complicated constructive and destructive interference patterns that are strongly dependent on the environmental parameters~\cite{wang2014review, DiNapoli1979,5422478}, conventional physics-based models require accurate environmental knowledge to make good predictions of the acoustic field. Accurate measurement of environmental parameters such as sediment properties, bathymetry, internal waves, suspended bubbles, surface wave spectra, etc. may be difficult or expensive in practice. Even in cases where such information is available, it may not always be straightforward to incorporate it into propagation models.}

\update{While the limitations in physics-based models pose a practical hurdle for many underwater applications, the emergence of data-driven machine learning~(ML) algorithms might offer a promising alternative. ML algorithms allow computers to automatically learn from data and perform certain tasks that were previously considered difficult~\cite{goodfellow2016machine, jordan2015machine}.} \update{We can model acoustic field using classical data-driven ML techniques} such as Gaussian process regression~(GPR)~\update{\cite{caviedes2021gaussian},\cite{kohlsche2019gaussian}} \update{and artificial} neural networks~(NN\update{s})~\update{\cite{lee2022predicting, mallik2022predicting}}, provided sufficient acoustic field measurements are available as training data for the models. The Gaussian process model is a probabilistic model widely used for regression and classification problems. GPR is capable of capturing relations between inputs and outputs through non-parametric Bayesian inference~\cite{bernardo1998regression,Rasmussen2004,seeger2004gaussian,murphy2022probabilistic}. Given a set of acoustic measurements and the corresponding measurement locations, GPR can interpolate and extrapolate acoustic predictions with uncertainty distributions at unvisited locations. The computational cost of the traditional GPR scales poorly with the \update{size of} training data, but sparse GPR models have been proposed to counter such a limitation~\cite{liu2020gaussian}. The availability of a large amount of accessible datasets has driven the rapid growth in the development of NN algorithms over the past years. \update{NNs are composed of interconnected neurons arranged in layers. NNs learn to identify complicated patterns and relationships in training data by properly adjusting weights and biases of all neurons~\cite{gurney2018introduction}.} The universal function approximation theorem establishes that multilayer feed-forward \update{NNs} have the capability to approximate any continuous function given a sufficient number of hidden units~\cite{hornik1989multilayer}. This suggests that a NN should be able to approximate the solution \update{of} the wave equation by learning appropriate weights \update{and biases} from the training data. 

Since data-driven approaches only require acoustic measurements for training, they eliminate the need to have full and accurate prior environmental knowledge. \update{Once the model is trained, evaluating 3D acoustic fields at unvisited locations is much faster than 2.5D numerical methods. These factors make data-driven methods a feasible alternative for predicting acoustic fields in cases where physics-based models fail.} However, the two key problems that limit their use in \update{practical} acoustic propagation modeling are the necessity of a large training dataset, and the inability to extrapolate well~\cite{2017}. The cost of acoustic data acquisition is inevitably high, as the ocean environment tends to be expensive to operate in. In this paper, we address these two problems by developing hybrid propagation modeling methods that not only learn from data, but also utilize the knowledge of the physics of acoustic propagation, without requiring full environmental knowledge. \update{We leverage the complementary strengths of physics-based propagation models and data-driven ML to develop a hybrid approach that utilizes available environmental knowledge, requires limited training data, and extrapolates well.}

\begin{figure}[t]
\centering
\includegraphics[width=0.6\textwidth]{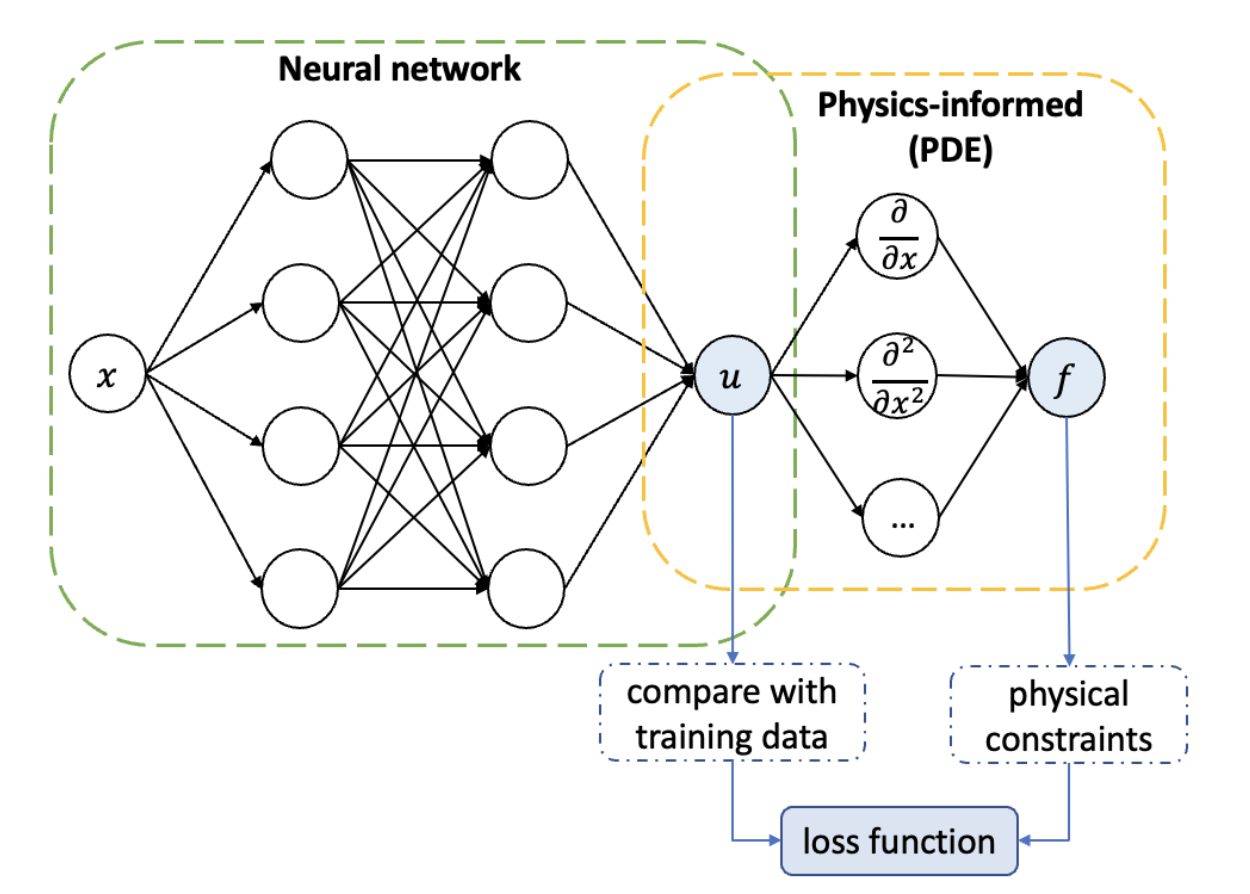}
\caption{The typical model structure of PINNs. \update{Node $u$ denotes the predicted output of the NN and node $f$ denotes the residual, measuring how well the predictions satisfy the underlying physical PDE. The residual term becomes 0 in the loss function when NN prediction fully satisfies the underlying physics.}} 
\label{fig:PINN_schematic}
\end{figure}

% Literature review of SciML, PINN:
The need to combine knowledge of physics with data-driven ML is not limited to ocean acoustic modeling, and is in fact the focus of an emerging field called \emph{scientific machine learning}~(SciML)~\cite{osti_1478744}. Researchers have explored synergistic ways that use scientific domain knowledge to aid data-driven ML~\cite{raissi2018hidden,swiler2020survey,willard2021integrating,DBLP:journals/corr/abs-2001-04385, read2019process, sun2019theory}. A popular SciML strategy, named physics-informed neural networks~(PINN\update{s}), imposes physics constraints in the form of partial differential equations~(PDE\update{s}) to act as a regularizer in the loss function of a NN~\cite{raissi2019physics}. Such an augmentation in the loss function helps to alleviate the problems of requiring large amounts of data, and the inability to extrapolate. A typical structure of PINN models is shown in Fig.~\ref{fig:PINN_schematic}. In the context of acoustic propagation, NNs can be informed by the \update{acoustic} wave equation to generate data-efficient solution approximations~\update{\cite{borrel2021physics, moseley2020solving,rasht2021physics}}. \update{However,} as far as we are aware, there are limited attempts at exploring the use of SciML for ocean modeling. One of the few works that assess the effectiveness of PINN\update{s} in solving simple ocean-related modeling problems is~\cite{de2021assessing}. Although there are many successful implementations showing that imposing physics knowledge improves model training, the authors of~\cite{de2021assessing} found no significant benefits to the training error by \update{having additional} physics-informed constraints in the loss function in their simulation studies. The authors of~\cite{de2021towards} \update{explored} the benefits of using PINN\update{s} to solve three ocean modeling-related PDEs, including the wave equation.

PINNs encode the physics as part of the loss function, and strike a balance between data-driven and physics-informed through hyper-parameters that control the weights of various terms in the loss function. \update{Hyper-parameter tuning is critical to the successful application of PINNs, but can often be difficult.} We take a different approach \update{to incorporate underlying physics in our models}. We design a class of ML algorithms where the physics is encoded in the structure of the algorithms. The functions these algorithms learn are automatically solutions to the acoustic wave equation. We give up the universal approximation property of NNs, and instead constrain our algorithms to only learn physically realistic functions. This constraint enables \update{learning} from very little data, and \update{extrapolation} beyond the region where the data \update{were} collected. Our approach is not only data-efficient, but also avoids the need for additional hyper-parameter tuning. Moreover, the algorithm is computationally simple and we are able to fully model 3D acoustic environments easily.

% key contributions:
\update{To the best of our knowledge, our previous preliminary works~\cite{kexin2021ocean,li2022physics} are the first attempts to explore the feasibility of modifying the structure of neurons in a NN in order to incorporate the underlying physics of ocean acoustics propagation. Ray models are computationally efficient and widely used in high-frequency underwater applications~\cite{hovem2013ray}. They can be seen as an intuitive way to interpret the wave equation solution as they track trajectories of a set of rays originating from the source as they propagate in oceans~\cite{Jensen2011}. In~\cite{kexin2021ocean}, we utilized ray models as the basis of our data-aided modeling. In this follow-up paper, we investigate this idea further and develop a general recipe\footnote{\update{In this paper, the term ``recipe" refers to a general approach that yields a class of physics-based data-aided propagation modeling algorithms.}} that targets solving forward problems in cases where there is a lack of environmental knowledge (but some acoustic measurements are available). Although inverse problems are not the focus of our work here, our modeling approach can also help solve some inverse problems as a by-product. The recipe presented in this paper is based on ray methods, but it can be extended to other physics-based modeling approaches (e.g. normal modes) as demonstrated in~\cite{li2022physics}.}

\update{The main contributions of this paper are summarized below:}
\begin{enumerate} 
\item We develop a recipe to generate physics-aided data-driven \update{forward acoustic propagation models that solve the acoustic wave equation in a desired number of dimensions (typically 1D, 2D or 3D)}.
\item The model does \update{not} require accurate environmental knowledge, but is able to utilize any available environmental knowledge.
\item The model does not require a large amount of training data, and has the ability to extrapolate beyond the region where data \update{were} collected.
\item The recipe supports composition, thus enabling us to combine purely data-driven ML models and physics-aided ML models into a single propagation model. For example, we may use a standard NN to model the reflection coefficient of the seabed, and combine it with a physics-aided model for the overall propagation modeling.
\item \update{Our framework brings interpretability to trained model parameters. This is particularly useful in inverse problems, where environmental information may be extracted from the trained model parameters.}
\item We demonstrate our \update{method} with several numerical experiments, and benchmark it against \update{two popular data-driven techniques -- GPR and deep neural network~(DNN)}. We also carry out a controlled experiment in a water tank to validate the performance of a generated propagation model.
\end{enumerate}

The rest of the paper is organized as follows. In Section~\ref{sec:method}, we introduce the modeling recipe and illustrate three \update{example formulations arising from the use of the recipe}. In Section~\ref{sec:simulation_studies}, we numerically demonstrate some use cases of the proposed framework for four common oceanic applications. This is followed by \update{an} experimental validation in a water tank environment in Section~\ref{sec:experiment}. The field estimation results are benchmarked against a GPR model and a DNN model. 
\update{In Section~\ref{sec:discussion}, we highlight the assumptions and limitations of our proposed recipe. We also discuss potential solutions and possible extensions.} In Section~\ref{sec:conclusion}, we summarize our findings \update{and conclude this paper}.

Some abbreviations and symbols used throughout the paper are listed in Table~\ref{tab:abbreviations} and Table~\ref{tab:symbol_list}.

\begin{table}[t]
\caption{Abbreviations used in the paper.}
\renewcommand{\arraystretch}{1.25} 
\centering
{\small
\begin{tabular}{ll}
\toprule \toprule
{Abbreviation}  & {Description}\\
\midrule
GPR & Gaussian Process Regression\\
ML & Machine Learning\\
NN & Neural Network \\
DNN & Deep Neural Network \\
SciML & Scientific Machine Learning\\
PINN & Physics-Informed Neural Network\\
PDE & Partial Differential Equation\\
AOI & Area Of Interest\\
CW & Continuous Wave\\
RBNN & Ray Basis Neural Network\\
RCNN & Reflection Coefficient Neural Network\\
ISM & Image Source Method\\
RMS & Root Mean Square\\
MATE & Mean Absolute Test Error\\
\bottomrule
\end{tabular}}
\label{tab:abbreviations}
\end{table}

\begin{table}[t]
\caption{Symbols used in the paper.}
\label{tab:symbol_list}
{\small
\begin{tabular}{llll}
\toprule \toprule
{Symbol}&{Description}&{Symbol}&{Description}\\
\midrule
$\boldsymbol{r}_s$ & Source location  & $n_\text{s} $ & Number of surface reflections\\
$f$ & Frequency & $e_\theta$ & Error to pre-calculated nominal azimuthal angle $\theta'$\\
$\boldsymbol{r}$ & Spatial coordinate & $e_\psi$ & Error to pre-calculated nominal elevation angle $\psi'$\\
$c$ & Sound speed  & $e_d$ & Error to pre-calculated nominal distance $d'$\\
$\omega$ & Angular velocity & $\boldsymbol{r}_\text{train}$ & Measurement locations in training dataset\\
$\bar p(\boldsymbol{r}) $& \update{Complex pressure amplitude} at location $\boldsymbol{r}$  & $\boldsymbol{y}_\text{train}$ & Acoustic measurements in training dataset\\
$k$ & Wavenumber & $\alpha$ & Penalty coefficient of arrival ray amplitude\\
$\boldsymbol{k}$ & Wave propagation vector & $\boldsymbol{\zeta}$ & Penalty coefficient of angular error\\
$A$ & Amplitude of a ray & $\beta$ & Penalty coefficient of distance error \\
$\phi$ & Phase of a ray  & $\eta$ & Penalty coefficient of reflection energy \\
${\theta}$ & Azimuthal angle & $\mathcal{T}$ & Trainable parameters in RBNN model \\
${\psi}$ & Elevation angle  & $\boldsymbol{R}$ & Trainable parameters in RCNN model\\
$\bar{d}$ & Propagation distance of a ray & $\epsilon$ & Reflection coefficient\\
$\boldsymbol{r}_o$ & Reference location  & $\kappa$ & Reflection phase shift \\
$\boldsymbol{s}$ & Image source location & $\gamma$ & Incident angle of a reflection\\
$d$ & Distance between $\boldsymbol{s}$ and $\boldsymbol{r}_o$ & $\Gamma$ & Complex reflection coefficient\\
$n_\text{ray}$ & Number of arrival rays & $l_{\text{a}}$ & Volume absorption loss term \\
$n_\text{ref}$ & Order of reflection & $l_{\text{rc}}$ & Reflection loss term \\
$n_\text{b} $ & Number of bottom reflections & $l_\text{g}$ & Geometric spreading loss term\\
\bottomrule
\end{tabular}}
\end{table}

$Notation$: Bold symbols and $[\cdot]$ denote vectors. Symbols in calligraphic font and $(\cdot)$ represent tuples. Sets are written as $\{\cdot\}$. We use the interval notation: $[a,b)= \{x \in \mathbb{R} | a \leqslant x < b\}$. $|c|$ denotes the magnitude of a complex number $c$. For vectors $\boldsymbol{a}$ and  $\boldsymbol{b}$, $\boldsymbol{a} \cdot \boldsymbol{b}$ is the dot product. $\rVert\boldsymbol{a}\rVert_1$ and $\rVert\boldsymbol{a}\rVert_2$ denote $L_1$-norm and $L_2$-norm of vector $\boldsymbol{a}$ respectively. The symbol $\equiv$ denotes equivalence, and the symbol $\nabla^2$ is the Laplace operator.

\section{Problem \& Solution Formulation}\label{sec:method}

\subsection{Ray-basis Neural Network Framework}
We consider an acoustic propagation modeling problem where we have limited environmental knowledge and a small amount of acoustic data collected within \update{an} area of interest~(AOI). We assume an acoustic source is located at position~$\boldsymbol{r}_s$ and \update{omni-directionally} transmits a continuous wave~(CW) signal at frequency $f$. The signal emitted from the source is scattered by the water surface and other boundaries as it propagates through the acoustic channel. The received signal $p(\boldsymbol{r})$ at location $\boldsymbol{r}$ is composed of a sum of multipath arrivals, each with associated intensity and arrival time. The constructive and destructive interference due to the multipath may lead to strong spatial variations in the acoustic field within AOI.
 
The acoustic wave equation determines the propagation of the acoustic energy from a source, and is expressed as~\cite{Jensen2011propagation}:
\begin{equation}
\frac{\partial^2 p} {\partial{t^2}} =c^2\nabla^2 p,
\label{eq:wave_equation}
\end{equation}
where $p$ represents acoustic pressure, $t$ denotes time and $c$ is the sound speed \update{in water\footnote{\update{In a general formulation, the sound speed $c$ may vary with position and depth. While the method in this paper can be extended for application in such cases, in this paper, we mostly restrict our discussion to the case where $c$ is constant in the AOI for field estimation.}}}. A solution to~\eqref{eq:wave_equation} can be written as:
\begin{equation}
p(\boldsymbol{r}, t) =\bar{p}(\boldsymbol{r}) e^{i\omega t},
\label{eq:general_sol}
\end{equation}
where $\boldsymbol{r}$ is the spatial coordinate, $\bar{p}(\boldsymbol{r})$ represents complex pressure amplitude, and $\omega = 2\pi f$ denotes angular frequency. Substituting~\eqref{eq:general_sol} back into~\eqref{eq:wave_equation}, and rearranging, we get the Helmholtz equation~\cite{Jensen2011propagation}:
\begin{equation}
k^2\bar{p}(\boldsymbol{r})  + \nabla^2 \bar{p}(\boldsymbol{r}) = 0,
\label{eq:hel_eqn}
\end{equation}
where $k = \frac{\omega}{c}$ is called the wavenumber. 
Equation~\eqref{eq:hel_eqn} can be solved by:
\begin{equation}
\bar{p}(\boldsymbol{r}) =Ae^{i\phi}e^{i {\boldsymbol{k}}\boldsymbol{\cdot} {\boldsymbol{r}}},
\label{eq:solution1}
\end{equation}
where $A$ and $\phi$ represent amplitude and phase of a \update{plane} wave, and $\boldsymbol{k}$ is \update{the} wave propagation vector \update{pointing normal to the wavefront, such that $\| \boldsymbol{k} \|_2= k$}.

Any function of the form~\eqref{eq:solution1} solves the wave equation. \update{Due to the linearity of the wave equation,} the superposition of $n_\text{ray}$ such \update{plane-wave solutions (each associated with a \emph{ray})} must also be a solution to the wave equation. Thus, the field at a location $\boldsymbol{r}$ can be expressed as the sum of terms given by~\eqref{eq:solution1}:
\begin{equation}
\bar{p}(\boldsymbol{r}) = \sum_{m=1}^{n_\text{ray}} {A}_m e^{i{\phi}_m}  e^{i {\boldsymbol{k}_m} \boldsymbol{\cdot} {\boldsymbol{r}}},
\label{eq:sum_sol}
\end{equation}
where ${{A}_m}$ denotes the amplitude of $m^{\text{th}}$ arrival, ${\phi}_m$ refers to the corresponding phase term, and $\boldsymbol{k}_m = k \hat{\boldsymbol{k}}_m$ for some unit vector $\hat{\boldsymbol{k}}_m$.

\begin{figure}[t]
\centering
\includegraphics[width=0.7\textwidth]{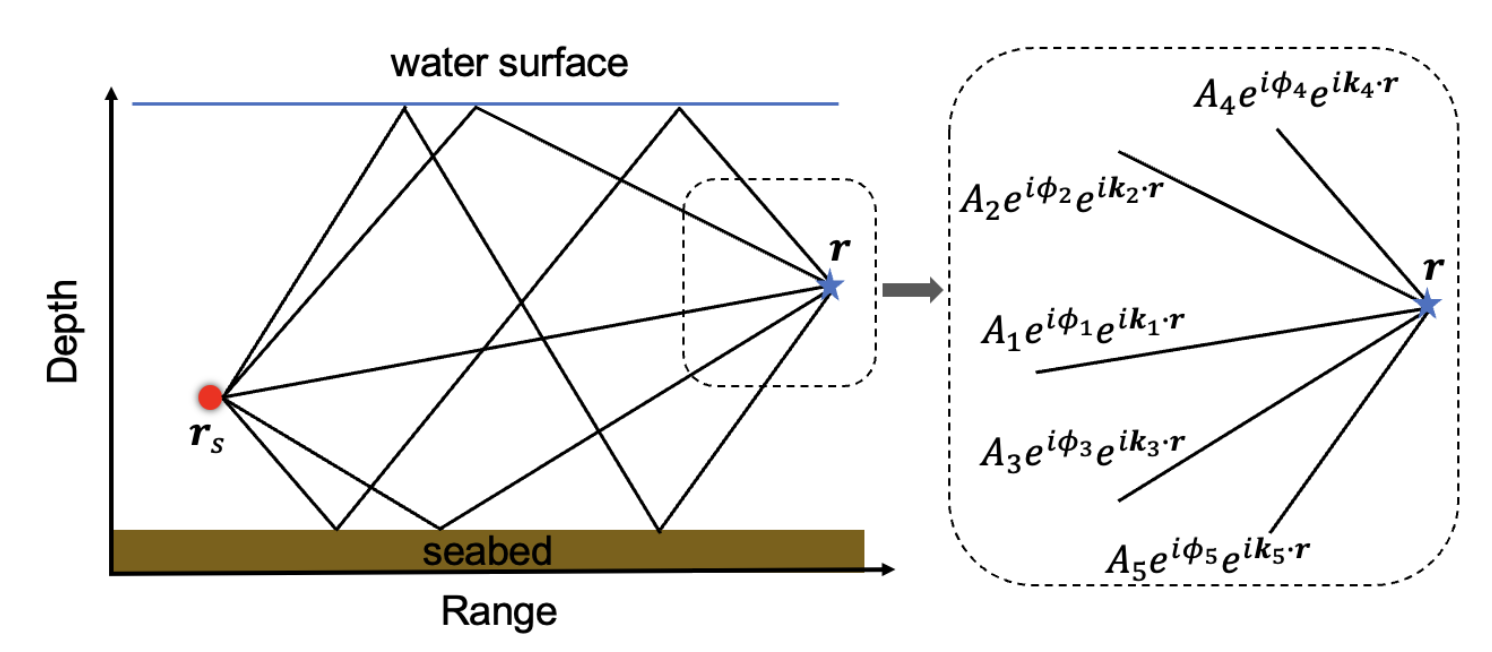}
\caption{An example showing the superposition of 5 multipath arrivals at a receiver location $\boldsymbol{r}$.}
\label{fig:superposition}
\end{figure}

This is the well-known ray solution to the acoustic wave equation~\cite{Jensen2011}, with ${A}_m e^{i{\phi}_m}$ being the complex amplitude of the $m^\text{th}$ ray and $\hat{\boldsymbol{k}}_m$ being the direction of travel of that ray. Fig.~\ref{fig:superposition} elaborates \update{on} the \update{intuitve interpretation} behind~\eqref{eq:sum_sol} from a receiver point of view. The acoustic field at a receiver location $\boldsymbol{r}$ can be visualized as the superposition of $n_\text{ray}$ multipath arrivals. Conventional ray models determine ${{A}_m}$, ${\phi}_m$ and $\boldsymbol{k}_m$ for all $m$, given detailed environmental knowledge. It is not generally possible to compute ${{A}_m}$, ${\phi}_m$ and $\boldsymbol{k}_m$ if partial or no environmental knowledge is available. Fortunately, ML provides us the necessary tools to learn unknown parameters or functions from data. We can think of~\eqref{eq:sum_sol} as a function to be modeled using a specialized NN with each term in the summation playing the role of a neuron\footnote{\update{The term \emph{neuron}, while originally inspired by biological neurons, is now used in the ML literature to denote a trainable unit of computation in the NN. Layers of various types of neurons (e.g. fully connected layers, convolution layers, etc.) are commonly combined together to form a NN. Each term in~\eqref{eq:sum_sol} is a trainable computational unit and hence plays the role of a neuron, and the summation of all terms plays the role of a layer of neurons in a NN.}} (with parameters closely related to ${{A}_m}$, ${\phi}_m$ and $\boldsymbol{k}_m$). The values of the parameters can be \update{learned} from data using a generalized \emph{backpropagation} algorithm~\cite{hecht1992theory},\update{\cite{werbos1974beyond}} with automatic differentiation~\cite{baydin2018automatic} applied to this NN. \update{Various optimization algorithms such as ADAM~\cite{Kingma2015AdamAM}, SGD~\cite{ketkar2017stochastic}, L-BFGS~\cite{berahas2016multi}, etc., commonly used in ML, may be utilized for \emph{training} (learning of the model parameters from data)}.

\begin{figure}[t] 
\centering
\includegraphics[width=0.6\textwidth]{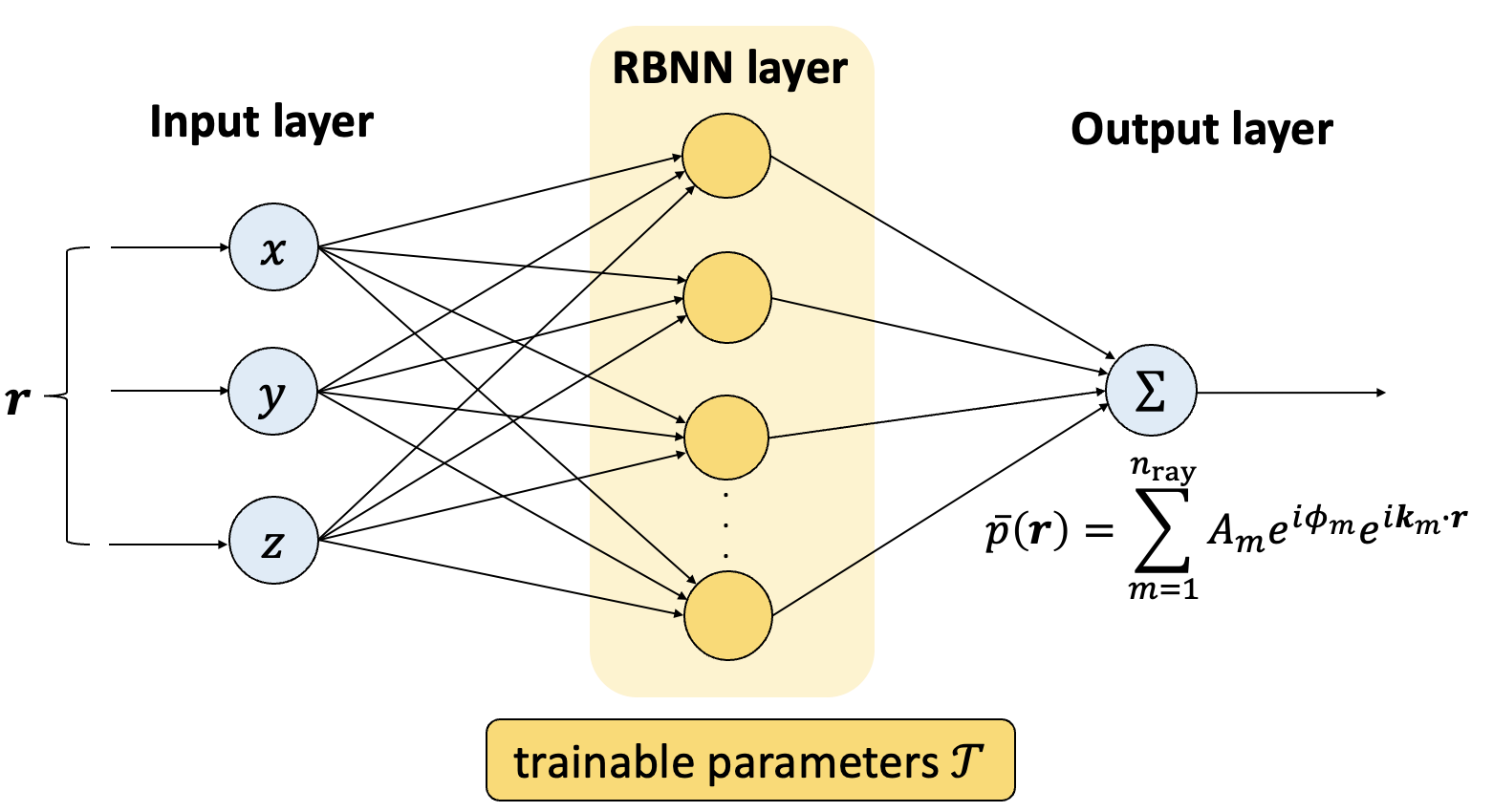}
\caption{ \update{The computation graph for computing total pressure amplitude as the sum of contributions from each ray as shown in~\eqref{eq:sum_sol} using the proposed RBNN framework. The trainable parameters~$\mathcal{T}$ include all unknown RBNN model parameters yet to learn from data.}}
\label{fig:RBNN_structure}
\end{figure}

The functions that this NN can learn are guaranteed to solve the wave equation~\eqref{eq:wave_equation} by construction, hence incorporating the acoustic domain knowledge in the structure of the NN. We term this specialized NN as a \textit{ray basis neural network}~(RBNN), as the neurons in the network can be interpreted as acoustic \update{plane waves} arriving at a given receiver location as shown in Fig.~\ref{fig:RBNN_structure}.

%%% RBNN diagram:

It is worth noting that our model structure is determined \update{by the} governing physics and is conceptually different from a conventional PINN. Instead of imposing physical constraints using the loss function of a standard NN, we embed the constraints in the structure of the NN by making each neuron individually obey the governing physics. We then \update{use the same training strategies that standard NNs typically used to find the best-fitted} values of the unknown parameters of the resulting model by continuously minimizing a loss function that measures the error between the model output and observed data.
The dataset used in the model training stage is comprised of a set of measurement locations $\boldsymbol{r}_{\text{train}}$ and corresponding acoustic field measurements $\boldsymbol{y}_{\text{train}}$.
Furthermore, one can construct more sophisticated NNs with a mix of standard neurons and RBNN-neurons, as we shall show later. Such NNs can be useful in solving problems with partial environmental knowledge.

The formulation presented above forms a basic recipe to model high-frequency acoustic propagation using \update{SciML}. The exact calculations of ${{A}_m}$, ${\phi}_m$ and $\boldsymbol{k}_m$ of each ray are application-dependent, since some of these terms may be calculated based on environmental knowledge, and others determined from parameters learned from data. In Sections~\ref{sec:plane} and~\ref{sec:spherical}, we apply the recipe to generate models to handle three different application scenarios: plane wave (far-field propagation), spherical wave (near-field propagation) without knowledge of geometry, and spherical wave with knowledge of geometry. In each scenario, the exact details of RBNN-neurons change, but the overall RBNN structure and the training process \update{remain} the same.

\subsection{Plane wave RBNN} \label{sec:plane}

In the far-field of a point source, a ray arrival can be well approximated by a planar wavefront. So, if the AOI is sufficiently far from the source, we can use a plane wave formulation for the unknowns in~\eqref{eq:sum_sol}. This formulation does not require any prior environmental knowledge, and is particularly helpful to model practical scenarios where the environment is largely unknown. The unknown terms ${{A}_m}$ and ${\phi}_m$ are treated as unknown model parameters to be determined during training. If the sound speed or frequency is unknown, $k$ may also be treated as an unknown parameter. The unit vector $\hat{\boldsymbol{k}}_m$ is parametrized in terms of azimuthal angle $\theta_m$ and elevation angle $\psi_m$:
\begin{equation}
\hat{\boldsymbol{k}}_m = \begin{bmatrix}
  \cos(\theta_m) \:\sin(\psi_m)\\
  \sin(\theta_m) \: \sin(\psi_m)\\
  \cos(\psi_m)\end{bmatrix}.
\end{equation}
The set of trainable RBNN model parameters in the plane wave formulation therefore is:
\begin{equation}
\mathcal{T}_\text{p} \equiv \update{(}\boldsymbol{A}, \boldsymbol{\phi}, k, \boldsymbol{\theta}, \boldsymbol{\psi} \update{)},
\label{eq:plane_para}
\end{equation}
where  $\boldsymbol{A} = \left [A_1, A_2, \dots, A_{n_{\text{ray}}} \right ]$, $\boldsymbol{\phi} = \left [ \phi_1, \phi_2, \dots, \phi_{n_{\text{ray}}} \right ]$, $\boldsymbol{\theta} = \left [ \theta_1, \theta_2,\dots,\theta_{n_{\text{ray}}} \right ]$ and $\boldsymbol{\psi} = \left [\psi_1, \psi_2,\dots,\psi_{n_{\text{ray}}}\right ]$. The \update{complex} pressure amplitude predicted at location $\boldsymbol{r}$ can be expressed as \update{\eqref{eq:sum_sol}.}

Since we do not assume detailed environmental information, the number of rays $n_\text{ray}$ is unknown, but a conservative upper bound can often be estimated. Nevertheless, we find it better to think of $n_\text{ray}$ as a model hyper-parameter to be tuned during training, with the tuning guided by an estimate, if available. Due to the strongly non-linear effect of parameters $\theta_m$ and $\psi_m$, the \update{RBNN} may get trapped in local minima or saddle points during training if $n_\text{ray}$ is small. A large $n_\text{ray}$ and uniformly distributed random initialization of $\boldsymbol{\theta}$ and $\boldsymbol{\psi}$ ensure better convergence, but create potential for overfitting. A $L_1$-norm regularization on parameters $\boldsymbol{A}$ encourages sparsity, i.e., a trained model with only a small number of rays, and therefore avoids overfitting.

The loss function we minimize during the training is therefore the sum-square difference in predicted and measured pressure amplitudes at given receiver locations, combined with the $L_1$-norm regularization term to encourage sparsity:
\begin{equation}
\update{L_{\text{p}}(\boldsymbol{r}, y; \mathcal{T}_\text{p})} = \left(|\bar{p}(\boldsymbol{r})| - y \right)^2 + \alpha \left \lVert \boldsymbol{A} \right \rVert_1,
\label{eq:p_loss_func}
\end{equation}
where $y$ is the observed pressure amplitude at location $\boldsymbol{r}$, and $\alpha$ is a hyper-parameter that controls the regularization. While we write~\eqref{eq:p_loss_func} for a single training data point, it is usually summed over a training mini-batch as per the standard practice in ML~\cite{li2014efficient}. During validation and model evaluation, $\alpha$ \update{in~\eqref{eq:p_loss_func}} is set to 0 \update{to only calculate the prediction error term.}

\subsection{Spherical wave RBNN} \label{sec:spherical}
The acoustic propagation near a point source is best modeled using spherical waves. In a typical ocean environment, there are three key factors that contribute to the overall transmission loss: geometric spreading loss $l_\text{g}$, volume absorption loss $l_{\text{a}}$ and reflection loss $l_{\text{rc}}$ (net effect from all reflecting boundaries)~\update{\cite{Jensen2011fun}}. In contrast to the plane wave formulation, the amplitude $\bar{A}$ and phase $\bar{\phi}$ of an arrival ray in our spherical wave formulation are functions of both source location $\boldsymbol{r}_\text{s}$ and receiver location $\boldsymbol{r}$. Therefore, \eqref{eq:sum_sol} is re-written as~\update{\cite{Kistovich2020,Jensen2011propagation,morse1986theoretical}}: 

\begin{equation}
\begin{split}
\bar{p}(\boldsymbol{r}) &= \sum_{m=1}^{n_\text{ray}} {\bar{A}_m(\boldsymbol{r}_\text{s},\boldsymbol{r}) e^{i{\bar{\phi}}_m(\boldsymbol{r}_\text{s},\boldsymbol{r})}},
\end{split}
\label{eq:sph_sum}
\end{equation}
where

\vspace{-0.5em}
\begin{subequations}
\begin{align}
\bar{A}_m(\boldsymbol{r}_\text{s}, \boldsymbol{r}) &= l^{m}_\text{g}(\boldsymbol{r}_\text{s}, \boldsymbol{r})l_{\text{rc}}^{m}(\boldsymbol{r}_\text{s}, \boldsymbol{r})l^m_{\text{a}}(\boldsymbol{r}_\text{s}, \boldsymbol{r}), \\
\bar{\phi}_m(\boldsymbol{r}_\text{s}, \boldsymbol{r}) &= \phi^m_\text{rc}(\boldsymbol{r}_\text{s}, \boldsymbol{r}) + k \bar{d}_m(\boldsymbol{r}_\text{s}, \boldsymbol{r}).
\end{align}
\end{subequations}
Here, $\phi^m_\text{rc}(\boldsymbol{r}_\text{s}, \boldsymbol{r})$ is the overall reflection phase shift along the trajectory of the $m^\text{th}$ ray, and $k \bar{d}_m(\boldsymbol{r}_\text{s}, \boldsymbol{r})$ corresponds to the phase change for a propagation distance of $\bar{d}_m(\boldsymbol{r}_\text{s}, \boldsymbol{r})$. The 3D spherical geometric spreading loss is~\cite{Jensen2011fun}:
\begin{equation}
l^{m}_\text{g}(\boldsymbol{r}_\text{s},\boldsymbol{r}) = \frac{1}{\bar{d}_{m}(\boldsymbol{r}_\text{s},\boldsymbol{r})}.
\end{equation}
The volume absorption loss generally depends on the operating frequency, propagation distance and characteristics of the propagating medium. The widely used simplified expression of the attenuation per unit distance due to volume absorption is given in \cite{fisher1977sound}. The attenuation and phase shift when sound interacts with scattering boundaries (e.g. seabed) can also be calculated if we know the angle of interaction and the properties and structure of the boundary~\cite{Brekhovskikh2003}.

The spherical wave formulation can incorporate varying degrees of environmental knowledge. The model parameters involved in the field prediction can be found through either data-driven learning strategies or numerical calculations, \update{depending} on the environmental knowledge provided. We next illustrate two examples that correspond to the scenarios with and without knowledge of the channel geometry:

\subsubsection{Without knowledge of channel geometry}
\begin{figure}[t]
\centering
\includegraphics[width=0.4\textwidth]{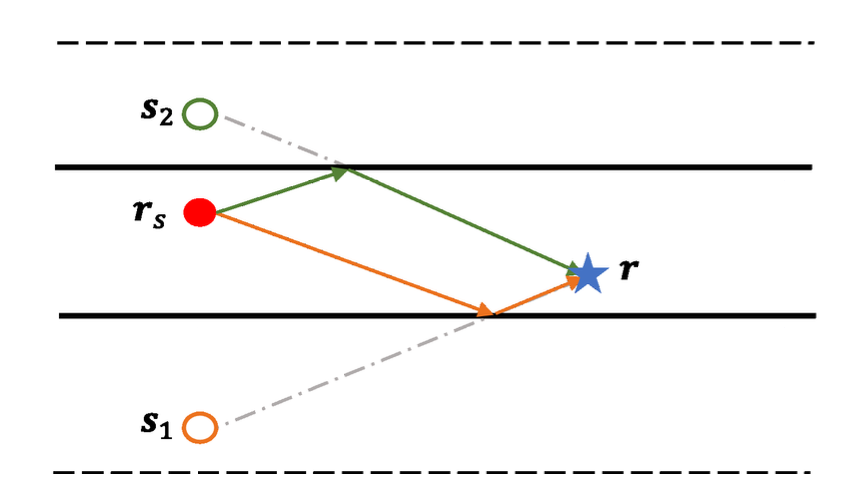}
\caption{An illustration showing two image sources, \update{$\boldsymbol{s_1}$ and $\boldsymbol{s_2}$}, corresponding to two reflected ray paths between the source \update{$\boldsymbol{r_s}$} and the receiver \update{$\boldsymbol{r}$}.}
\label{fig:reflection_order}
\end{figure}

For the scenarios where the channel geometry is largely unknown, the trajectories of rays from the source to the receiver are unknown. However, applying the image source method (ISM)~\cite{allen1979image} to the problem, we can replace the unknown source location and channel geometry by a set of unknown image sources as illustrated in Fig.~\ref{fig:reflection_order}. The problem then reduces to finding the parameters of the unknown image sources to match with the training data.

Let $\boldsymbol{r}_o$ be an arbitrary reference position within the AOI. We can parametrize each images source by a pressure amplitude $A_m$, phase $\phi_m$, a direction vector (corresponding to azimuthal angle $\theta_m$ and elevation angle $\psi_m$) and distance $d_m$ from this reference position. In the case of an isovelocity environment, the pressure amplitude at a receiver location $\boldsymbol{r}$ is then given by:
\begin{equation}
\update{\bar{p}(\boldsymbol{r})} = \sum_{m=1}^{n_\text{ray}} A_m{\frac{l_{\text{a}}(\left\lVert\boldsymbol{s}_m-\boldsymbol{r}\right\lVert_2) }{\left\lVert\boldsymbol{s}_m-\boldsymbol{r}\right\lVert_2}e^{i(\phi_m+ k \left\lVert\boldsymbol{s}_m-\boldsymbol{r}\right\lVert_2)}},
\label{eq:para_sph_without_geometry}
\end{equation}
where:
\begin{equation}
\boldsymbol{s}_m =\boldsymbol{r}_o -  d_m\begin{bmatrix}
\cos(\theta_m) \:\sin(\psi_m) \\
\sin(\theta_m) \: \sin(\psi_m) \\
\cos(\psi_m)
\end{bmatrix}
\label{eq:find_img}
\end{equation}
and $l_\text{a}(\cdot)$ is attenuation due to volume absorption as given in\cite{fisher1977sound}. 
The complete set of trainable parameters for this model are:
\begin{equation}
\mathcal{T}_\text{s} \equiv \left (k, \boldsymbol{\theta},\boldsymbol{\psi} ,\boldsymbol{d}, \boldsymbol{A}, \boldsymbol{\phi} \right ),
\label{eq:para_sph_without_geometry_ISM}
\end{equation}
\update{where $\boldsymbol{d} = \left [d_1, d_2, \dots, d_{n_{\text{ray}}} \right ]$}. The loss function to be minimized is identical to~\eqref{eq:p_loss_func}.

\subsubsection{With knowledge of channel geometry}\label{sec:sph_with_geo}
If the channel geometry and associated reflecting boundaries are partially or completely known, we can incorporate available knowledge \update{into} our model. To illustrate the idea, let us assume that we know the source location and the channel geometry. We also assume that the sea surface is modeled well as a pressure-release boundary, but that we do not know the reflection coefficient for the seabed.

\begin{figure}[t] 
\centering
\includegraphics[width=0.8\textwidth]{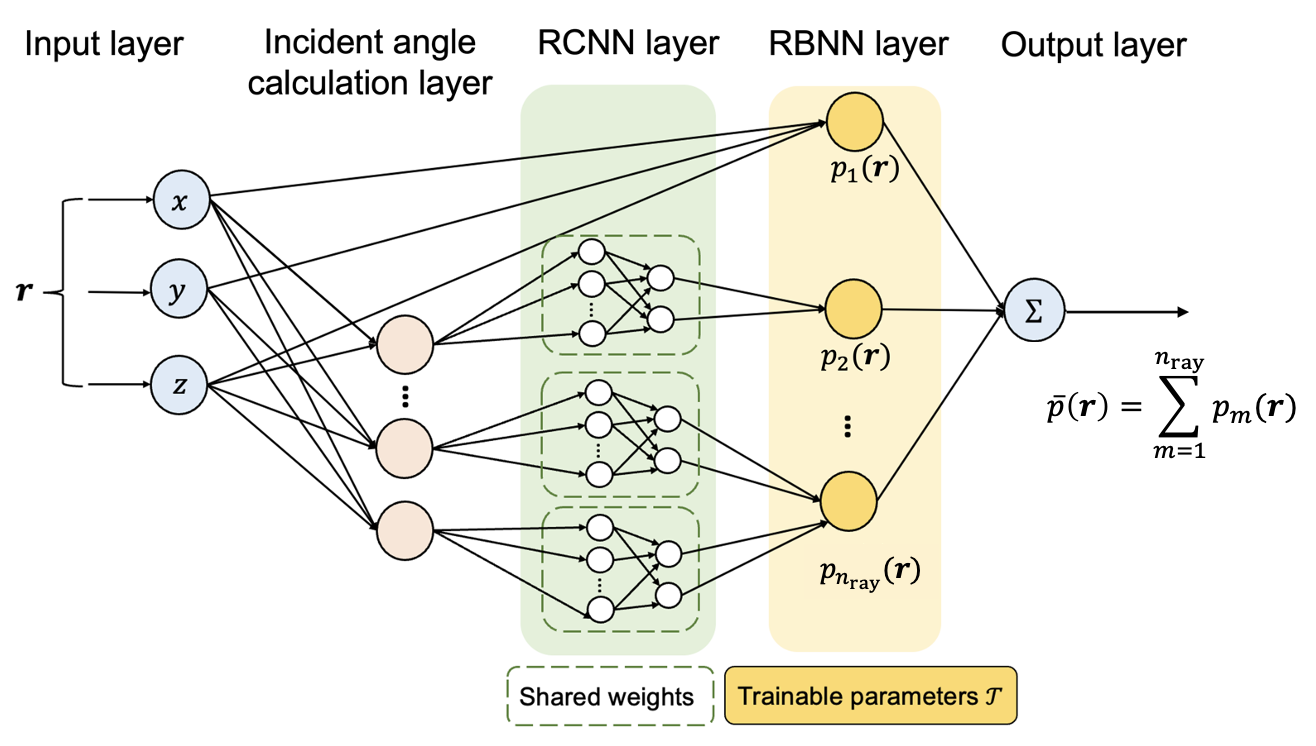}
\caption{\update{The computational graph for computing total pressure amplitude as the sum of contributions from each spherical ray as shown in~\eqref{eq:sph_amplitude}. We insert an additional RCNN layer to learn the unknown reflection model.}
\label{fig:RBNN_RCNN_structure}}
\end{figure}

Given the source location and channel geometry, we can compute the incidence angle $\gamma_m$ for each ray at the seabed. The reflection coefficient of the seabed is an unknown function of the incidence angle, and may be modeled using a simple 1-input 2-output (magnitude and phase) feedforward NN with a single hidden layer. We call this NN as the \textit{reflection coefficient neural network}~(RCNN). The same reflection coefficient function applies to all rays incident on the seabed, and hence the RCNN weights are shared across all the rays. The RCNN is implemented as an additional layer in the RBNN framework with shared weights, as illustrated in Fig.~\ref{fig:RBNN_RCNN_structure}.

A ray may experience more than one reflections at the seabed. The overall reflection coefficient for $m^\text{th}$ arrival ray is:
\begin{equation}
l_{\text{rc}}^{m}(\boldsymbol{r}) =  \prod_{i=1}^{n^{m}_{\text{b}}}\text{RCNN}_\epsilon(\gamma^{m}_i(\boldsymbol{r})),
\label{eq:rc}
\end{equation}
where $n^{m}_{\text{b}}$ is the number of seabed reflections for ray $m$, $\gamma^m_i(\cdot)$ is the incidence angle for reflection $i$, and $\text{RCNN}_\epsilon(\cdot)$ is the predicted reflection coefficient magnitude from the RCNN. The corresponding cumulative phase shift is:
\begin{equation}
\phi_{\text{rc}}^{m}(\boldsymbol{r}) =  n^{m}_{\text{s}} \pi + \sum_{i=1}^{n^{m}_{\text{b}}} \text{RCNN}_\kappa(\gamma^{m}_i(\boldsymbol{r})) ,
\label{eq:rc_phase}
\end{equation}
where $\text{RCNN}_\kappa(\cdot)$ is the reflection phase shift predicted by the RCNN, and $n^{m}_{\text{s}}$ is the number of surface reflections for ray $m$. The phase change $n^{m}_{\text{s}} \pi$ is due to the pressure-release boundary assumption, and can easily be replaced by a more sophisticated surface reflection model, if desired. 

As in the previous section, we choose to apply the ISM to replace the source with multiple image sources. This allows us to work with approximate knowledge of channel geometry and learn the exact locations of the image sources from data, as we illustrate later in this section. The resultant pressure amplitude can be expressed as:
\begin{equation}
\update{\bar{p}(\boldsymbol{r})} = \sum_{m=1}^{n_\text{ray}} {\frac{l_{\text{rc}}^{m}(\boldsymbol{r}) l_{\text{a}}(\left\lVert\boldsymbol{s}_m-\boldsymbol{r}\right\lVert_2) }{\left\lVert\boldsymbol{s}_m-\boldsymbol{r}\right\lVert_2} e^{i(\phi^m_\text{rc}(\boldsymbol{r})+k\left\lVert\boldsymbol{s}_m-\boldsymbol{r}\right\lVert_2)} },
\label{eq:sph_amplitude}
\end{equation}
where the ray trajectory necessary for the evaluation of $l_{\text{rc}}^{m}(\cdot)$ can be computed by geometric ray tracing. Since we are using a ray tracing model, \update{it may be possible to extend the algorithm to} non-isovelocity sound speed profile\update{s} by changing the \update{E}uclidean distances term\update{s} $\left\lVert\boldsymbol{s}_m-\boldsymbol{r}\right\lVert_2$ in \eqref{eq:sph_amplitude} to actual propagation distances along the curved ray paths.

The overall computation graph for~\eqref{eq:sph_amplitude} can be viewed as a NN with geometric ray tracer, RCNN layer and RBNN layer as shown in Fig.~\ref{fig:RBNN_RCNN_structure}. The set of trainable RBNN parameters in this model are:
\begin{equation}
\mathcal{T}_\text{sg}  \equiv \left (k, \boldsymbol{\theta}, \boldsymbol{\psi} ,\boldsymbol{d} , \boldsymbol{R}\right),
\label{eq:para_all_space}
\end{equation}
where $\boldsymbol{R}$ represents all trainable parameters in the RCNN layer.

The search spaces for $\boldsymbol{\theta}$ and $\boldsymbol{\psi}$ span $[0, 2\pi)$, and for $\boldsymbol{d}$ spans $[ 0, \infty)$. The knowledge of geometry and the source location allows us to pre-calculate nominal arrival ray directions $\boldsymbol{\theta}'$, $\boldsymbol{\psi}'$ and propagation distances $\boldsymbol{d}'$ prior to the model training stage. The calculated nominal directions and distances may deviate from reality due to limited knowledge or measurement error. We model this with appropriate error terms $\boldsymbol{e}_{\theta}$, $\boldsymbol{e}_{\psi}$ and $\boldsymbol{e}_d$:

\vspace{-0.5cm}
\begin{subequations}
\begin{align}
\boldsymbol{\theta} &= \boldsymbol{\theta'} + \boldsymbol{e}_\theta,\\
\boldsymbol{\psi} &= \boldsymbol{\psi'} + \boldsymbol{e}_\psi,\\
\boldsymbol{d} &= \boldsymbol{d'} + \boldsymbol{e}_d.
\end{align}
\end{subequations}
We then replace the trainable parameters $\boldsymbol{\theta}$, $\boldsymbol{\psi}$, and $\boldsymbol{d}$ with the corresponding error terms, thus replacing~\eqref{eq:para_all_space} with:
\begin{equation}
\mathcal{T}_\text{sg} \equiv \left (k, \boldsymbol{e}_{\theta},\boldsymbol{e}_{\psi} ,\boldsymbol{e}_d , \boldsymbol{R}\right).
\label{eq:sph_nominal_set}
\end{equation}

The amount of error allowed in $\boldsymbol{e}_\theta$, $\boldsymbol{e}_\psi$ and $\boldsymbol{e}_d$ reflect how confident we are about our knowledge of the channel geometry and source location. We impose $L_2$-norm penalty terms in the loss function to constrain values of $\boldsymbol{e}_\theta$, $\boldsymbol{e}_\psi$ and $\boldsymbol{e}_d$ \update{learned} during the training process. We also add a harsh penalty term to ensure that \update{the upper bound of reflected energy learned} by the RCNN obeys energy conservation. The resulting loss function is:

\vspace{-0.5cm}
\begin{equation}
%\begin{multline}
L_{\text{sg}}(\boldsymbol{r},y; \mathcal{T}_\text{sg}) = \update{\left( |\bar{p}(\boldsymbol{r})| - y \right)}^2 + \left \lVert \boldsymbol{\zeta}\update{\sqrt{\boldsymbol{e}_\theta^2 +\boldsymbol{e}_\psi^2}} \right \rVert_2 + \beta \left \lVert \boldsymbol{e}_d \right \rVert_2 + \eta \max \left\{ 0, \int_0^{0.5\pi} \text{\update{RCNN}}_\epsilon(\gamma)^2 d\gamma -1 \right\},
\label{eq: spherical_loss_func}
\end{equation}
%\end{multline}
where $\boldsymbol{\zeta}$, $\beta$ and $\eta$ are hyper-parameters related to the three penalty terms. All elements in the hyper-parameters are set to 0 during the validation and model evaluation. In general, the image sources \update{corresponding} to higher-order reflections are assigned smaller penalty coefficients as angular errors are amplified with an increasing number of reflections.

\section{Simulation Studies}\label{sec:simulation_studies}

\begin{figure}[t] 
\centering
\includegraphics[width= 0.7\textwidth]{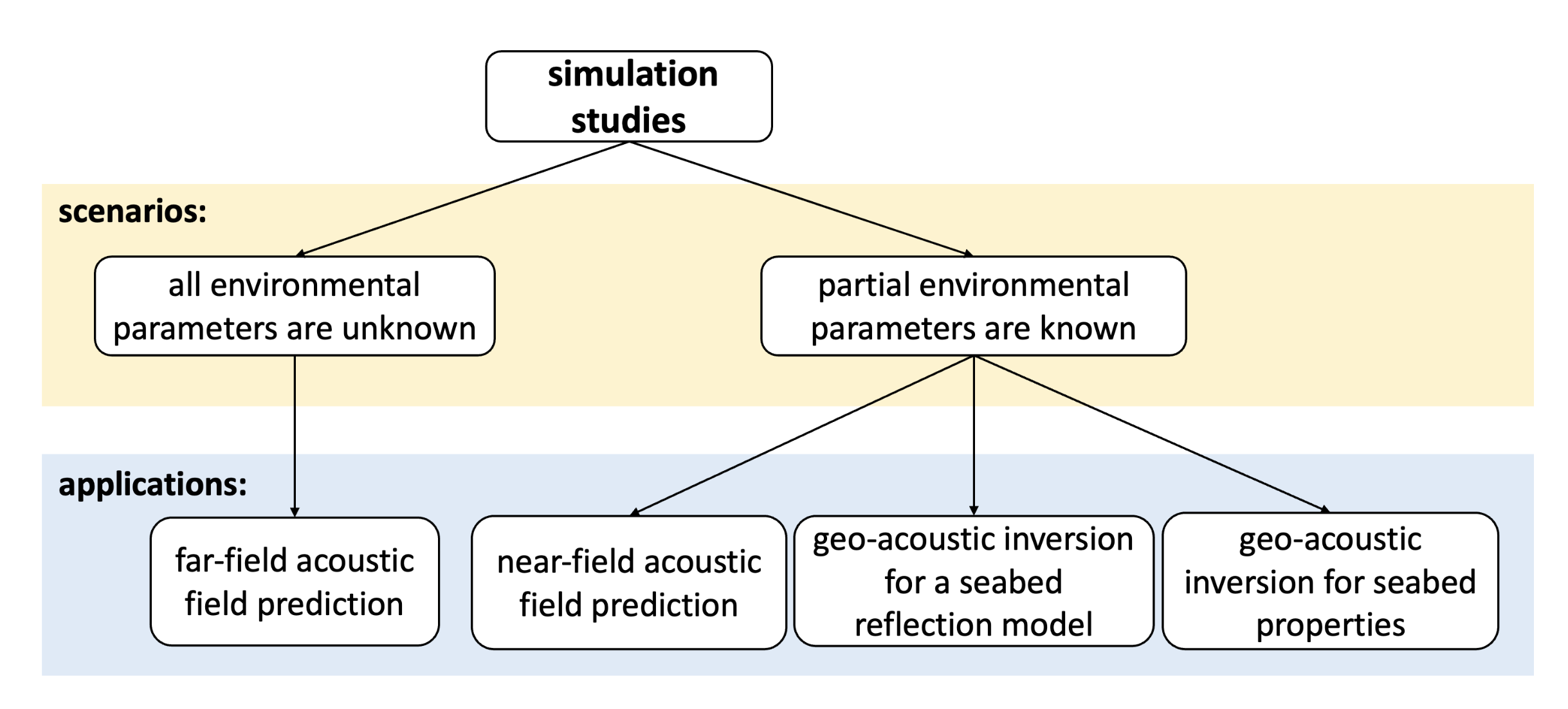}
\caption{Summary of the application scenarios demonstrated via simulation studies.} 
\label{fig:application}
\end{figure}

To study the effectiveness of our proposed method, we consider four common applications of ocean acoustic propagation models. These are summarized in Fig.~\ref{fig:application}.

All four applications considered use a profiling float equipped with a single hydrophone, collecting acoustic field measurements at a constant sampling rate. Such floats provide a cost-effective way of sparsely sampling acoustic \update{fields}. We assume that the profiling float can control its motion vertically, but not horizontally. The float freely drifts horizontally with ocean currents, thus following a zig-zag trajectory as it moves up and down through the water column. 

\subsection{Far-field acoustic field prediction} \label{sec:far_field_est}

\begin{figure*}[t] 
\centering
\includegraphics[width=1\textwidth]{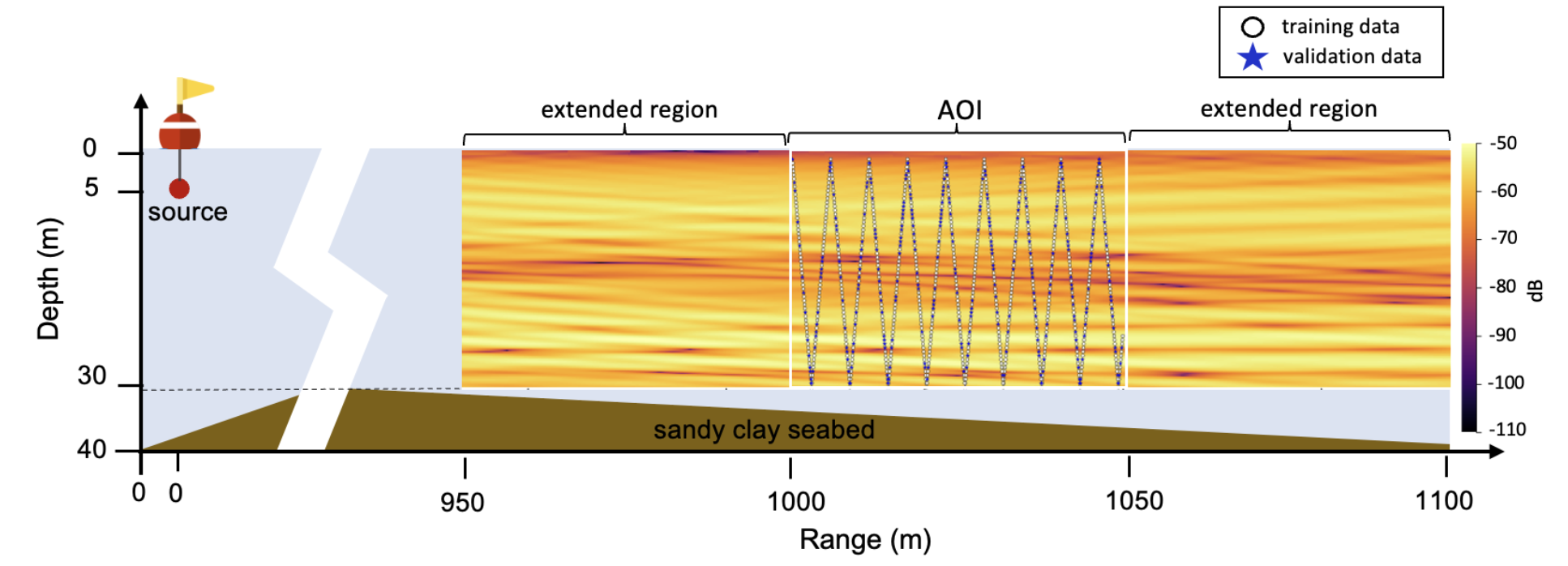}
\caption{\update{Simulated environment for the far-field acoustic field prediction application. The trajectory of the profiling float can be seen in terms of the training data points. The ground truth field pattern within the AOI is also shown.}}
\label{fig:plane_schematic}
\end{figure*}

The first application we shall consider is that of acoustic field prediction within an AOI at a long distance from an acoustic source. Suppose we have acoustic measurements from a profiling float along a zig-zag trajectory through the AOI \update{with a known sound speed}, but no other environmental knowledge. Conventional \update{physics-based} propagation models cannot be used for this application, as they require environmental knowledge as input. As discussed in Section~\ref{sec:plane}, RBNN parameters $\mathcal{T}_\text{p}$ can be \update{learned} without prior environmental knowledge, with acoustic field measurements as training data. \update{We formulate our RBNN model primarily based on \eqref{eq:sum_sol}}. With a sufficiently large distance between the acoustic source and the AOI, \update{we may expect that the curvature and path loss for all rays is similar. However, we find that modeling the curvature of the wavefront improves our field prediction accuracy. The set of trainable parameters is defined as:}
\begin{equation}
\update{\mathcal{T} \equiv \left ( \boldsymbol{A}, \boldsymbol{\phi}, \boldsymbol{\theta}, \boldsymbol{d} \right)},
\end{equation}
\update{where $\boldsymbol{A}$ and $\boldsymbol{\phi}$ represent amplitudes and initial phase of rays, $\boldsymbol{\theta}$ and $\boldsymbol{d}$ are used to model the wavefront curvatures of all rays.}
 \update{We use \eqref{eq:p_loss_func} as the loss function to train our RBNN model.}

\begin{table}[t]
\renewcommand{\arraystretch}{1.25} 
\caption{\label{tab:far_field_setup} Simulated environmental setup for the far-field acoustic field prediction application.}
\centering
\begin{threeparttable}
\begin{tabular}{ll}
\toprule \toprule
{Parameters} & {Value}\\
\midrule
Environmental model& 2D \\
Frequency & 10 kHz \\
Seabed & Sandy clay\tnote{1}\\
Bathymetry & Range-dependent\\
Source depth & 5 m\\
Sound speed & 1,541 m/s \\
Distance between source and AOI & \update{1,000}~m \\
Dimensions of AOI & 50~m~$\times$~30~m\\
Number of training data & \update{700} \\
Number of validation data& \update{300} \\
Number of test data \update{in AOI} & 601,601 \\
Number of rays in the RBNN layer & 60\\
\bottomrule
\end{tabular}
\begin{footnotesize}% or footnotesize, scriptsize, tiny, etc.
\begin{tablenotes}
  \item[1] \update{Sandy clay seabed is characterized by a relative density of 1.147, a relative sound speed of 0.9849 and a dimensionless seabed absorption coefficient of 0.00242~\cite{jackson1994apl}.}
 \end{tablenotes}
\end{footnotesize}
\end{threeparttable}
\end{table}

We simulate a profiling float performing \update{9} profiles through a 50~m $\times$ 30~m AOI at a distance of \update{1,000}~m from a 10~kHz source deployed at a depth of 5~m. The simulation setup is detailed in Table~\ref{tab:far_field_setup} and illustrated in Fig.~\ref{fig:plane_schematic}. A total of \update{1,000} acoustic field measurements are collected along the trajectory of the float, of which 70\% are used to train the \update{models}, and the remaining 30\% \update{are} used for validation. We wish to predict the acoustic field in the entire AOI.

We benchmark the field estimation performance of the RBNN against two popular data-driven techniques~--~GPR and DNN. We use a GPR with a composite kernel of a squared exponential isotropic kernel and a Mat$\acute{\text{e}}$rn $5/2$ ARD kernel\footnote{\update{Among the several commonly used kernels we tried, this type of composite kernel yields minimal validation error for our field estimation problems. We optimized the hyperparameters of this composite kernel using the L-BFGS technique based on training data.}}. The DNN model \update{takes location data as input and outputs corresponding pressure amplitude}. \update{It} has 3 fully-connected hidden layers with ReLU as \update{the} activate function. \update{We try various initial learning rates and choose the one that yields the smallest validation error.} The trainable model parameters in RBNN and DNN models are \update{randomly} initialized. We therefore carry out \update{20} Monte Carlo simulations for the RBNN and DNN models and present the results \update{of runs} with \update{the smallest} validation errors. \update{We implement early stopping~\cite{prechelt2012early} for RBNN and DNN model training to avoid overfitting.}

We use the Bellhop model~\cite{porter2011bellhop} to generate the \update{1,000} synthetic acoustic measurements along the profiler's trajectory for training and validation. The high-frequency source produces a complex interference pattern. To evaluate the field prediction performance in simulation, we generate a dense test dataset of 601,601 data points over a grid covering AOI, with a resolution of 0.05~m in depth and range. We also add a random position error of up to 0.1~m on each dimension of the measurement locations of the training and validation data to evaluate model robustness.

\begin{table}[t]
\renewcommand{\arraystretch}{1.25} 
\caption{\label{tab:field_model_complexity} Model complexity and RMS test error of the three models for field predictions within the AOI.}
\centering
\begin{threeparttable}
\begin{tabular}{llll}
\toprule\toprule
\multirow{2}{*}{{Method}}  & \update{Number of} & \multicolumn{2}{c}{{RMS test error}~(dB)} \\
\cline{3-4}
& \update{model parameters} &{Error-free data} & {Noisy data}\\
\midrule
RBNN\tnote{1} & 60 & \update{1.346} &  \update{1.678}  \\
GPR &  \update{1400\tnote{2}} & \update{1.783} & \update{1.637} \\
DNN & 6421 & \update{2.121 } & \update{2.180 } \\
\bottomrule
\end{tabular} 
\begin{footnotesize}\begin{tablenotes}
  \item[1] Plane-wave RBNN.
  \item[2] Dimensionality of each data point $\times$ training data size.
  \end{tablenotes}
\end{footnotesize}
\end{threeparttable}
\end{table}

\begin{figure*} 
\includegraphics[width=\textwidth]{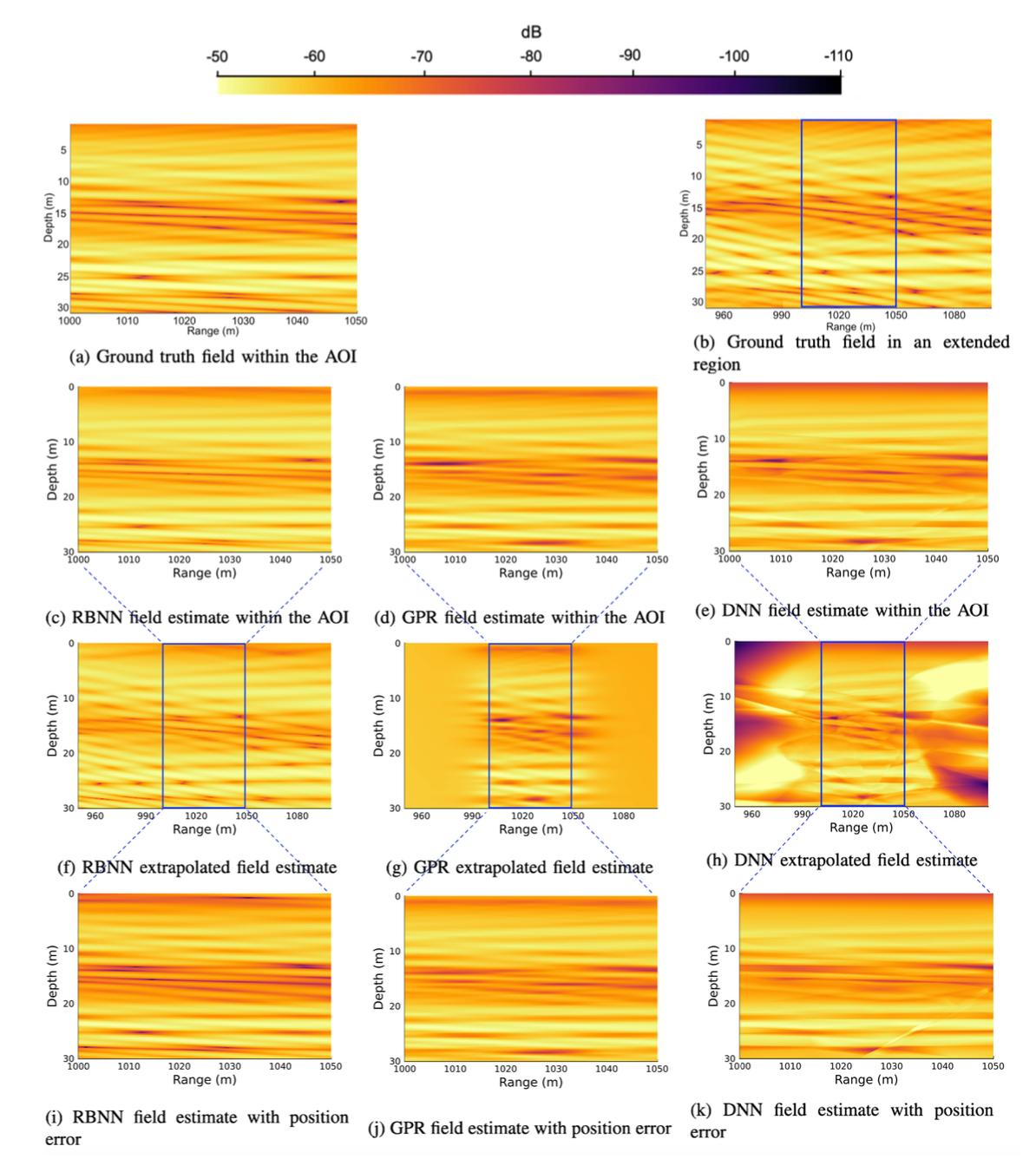}
\caption{The estimated field patterns for the far-field acoustic field prediction application. Panel (a) shows the ground truth field pattern within the AOI, while panel (b) shows the ground truth field within a 50~m extended area on both side\update{s} of the AOI. Panels (c)--(e) show the estimated fields by RBNN, GPR and DNN. Panels (f)--(h) show the corresponding extrapolated field by RBNN, GPR and DNN in the extended region. Panels (i)--(k) show the estimated field when the training data \update{have} positional errors.}
\label{fig:plane_field_est_res}
\end{figure*}

The root mean square~(RMS) test error and the model complexity (number of model parameters) of the three models are reported in Table~\ref{tab:field_model_complexity}. The acoustic field patterns estimated by the three approaches are shown in Fig.~\ref{fig:plane_field_est_res}. All three methods are able to learn key features of the acoustic field within the AOI. The RBNN model recovers most of the details in the AOI and extrapolates well in an \emph{extended region} beyond the AOI. The GPR learns the field pattern well within the AOI where training data \update{are} available, but fails to extrapolate the field in the extended region. The field pattern reconstructed by the DNN has the lowest fidelity among the three approaches. The extrapolated field by the DNN also deviates significantly from the ground truth. The extrapolated field patterns shown in Fig.~\ref{fig:plane_field_est_res} (f), (g) and (h) highlight the unique ability of the RBNN to not only interpolate well, but also to extrapolate.

The field estimation performance of the GPR and the DNN, as quantified by RMS test error, is not significantly affected by position errors in the dataset. On the other hand, the field estimation accuracy of the plane wave RBNN was found to be vulnerable to position errors. The qualitative field patterns seen in Fig.~\ref{fig:plane_field_est_res} (i), (j) and (k) show that the RBNN captures the overall field pattern best.

\subsection{Near-field acoustic field prediction} \label{sec:sph_field_est}

We next consider an acoustic field prediction application within an AOI from a less distant source assuming that the channel geometry is known. Acoustic measurements are collected along a zig-zag trajectory within AOI using a profiling float. Since some of the environmental parameters (e.g. seabed properties) are unknown, we cannot employ \update{physics-based} propagation models for field prediction. We can, however, use the spherical wave RBNN from Section~\ref{sec:sph_with_geo} with the knowledge of channel geometry. We can calculate the nominal arrival ray directions $\boldsymbol{\theta}'$, $\boldsymbol{\psi}'$ and propagation distances $\boldsymbol{d}'$ prior to the training. This significantly \update{reduces the training time and improves prediction accuracy}.

For this application, we use a simple spherical wave RBNN model based on~\eqref{eq:sph_amplitude}, without the RCNN layer. The overall effect of the reflections and absorption is modeled with a set of trainable parameters $\boldsymbol{A}$, associated with the set of rays. To model geometrical measurement errors, we also add error parameters for nominal direction and propagation distance. The set of trainable parameters therefore is:
\begin{equation}
\update{\mathcal{T} \equiv \left ( \boldsymbol{e}_{\theta},\boldsymbol{e}_{\psi} ,\boldsymbol{e}_d , \boldsymbol{A}, \boldsymbol{\phi} \right )}.
\end{equation}
\update{We use \eqref{eq: spherical_loss_func} as the loss function to train the trainable parameters and set $\eta$ in \eqref{eq: spherical_loss_func} to 0 as we do not have RCNN as part of the network.}

\begin{figure*}[t]
\includegraphics[width=\linewidth]{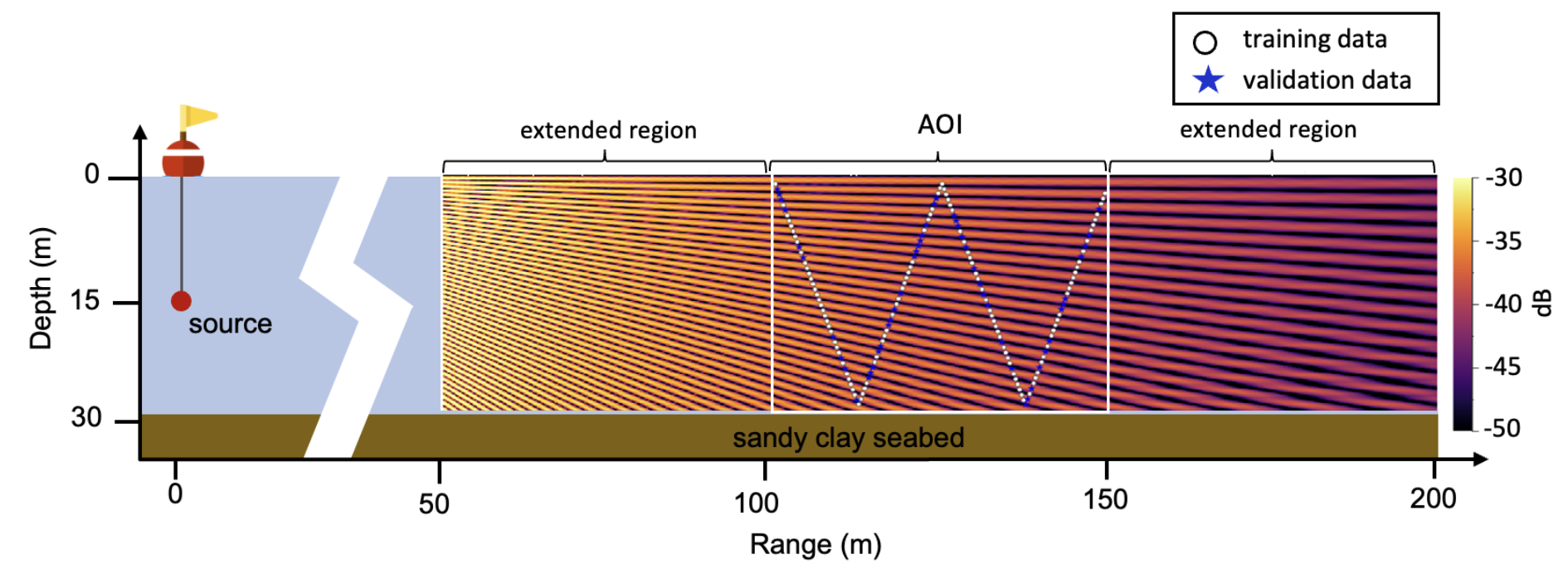}
\caption{\update{Simulated environment for the near-field acoustic field prediction application. The trajectory of the profiling float can be seen in terms of the training data points. The ground truth field pattern within the AOI is also shown.}}
\label{fig:sph_schemetic} 
\end{figure*}

\begin{figure}
\centering
\includegraphics[width=\textwidth]{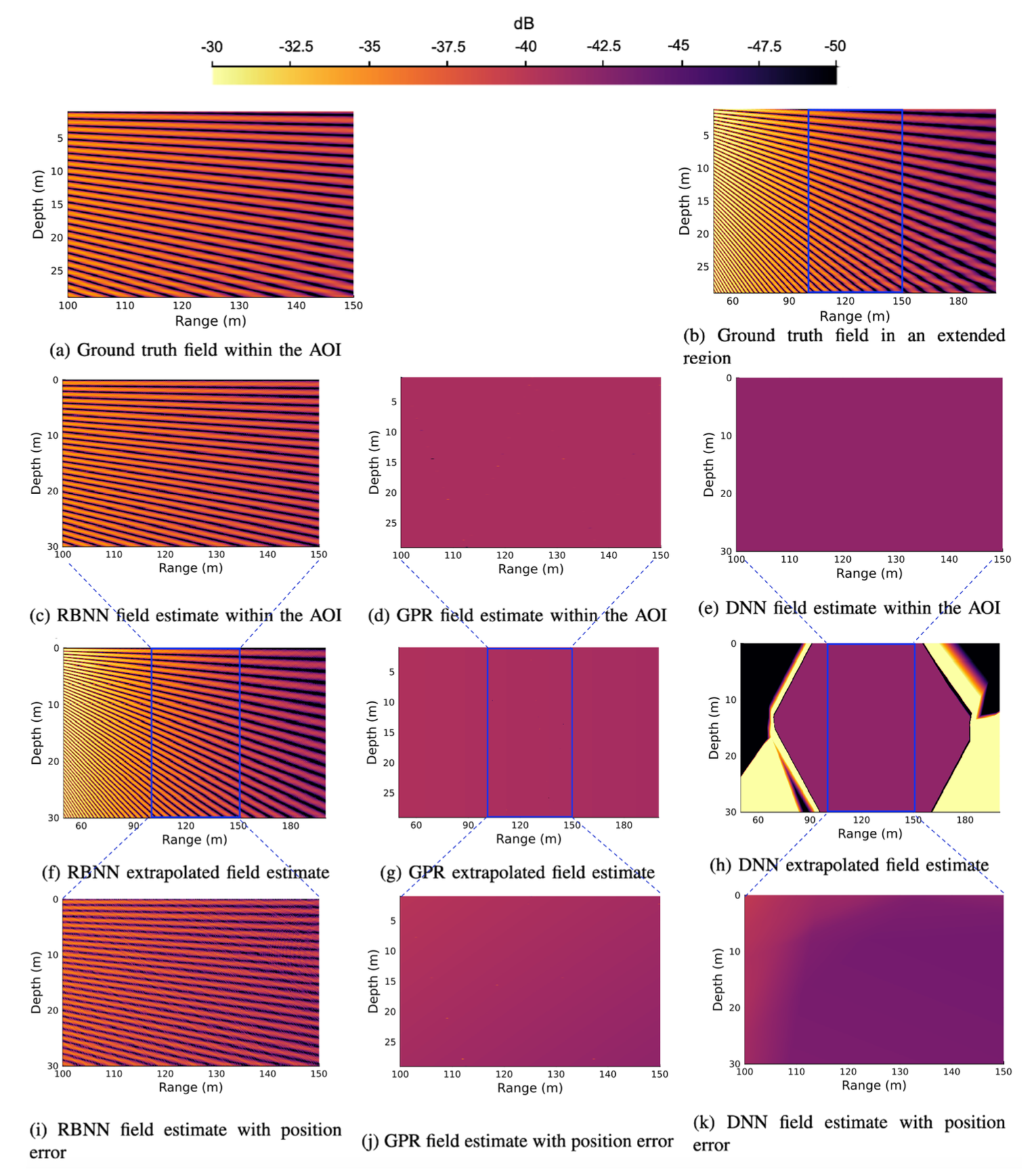}
\caption{The estimated field patterns for the near-field acoustic field prediction application. Panel (a) shows the ground truth field pattern within the AOI, while panel (b) shows the ground truth field within a 50~m extended area on both side\update{s} of the AOI. Panels (c)--(e) show the estimated fields by RBNN, GPR and DNN. Panels (f)--(h) show the corresponding extrapolated field by RBNN, GPR and DNN in the extended region. Panels (i)--(k) show the estimated field when the position and geometry measurements have random errors.}
\label{fig:sph_res}
\end{figure}

\begin{table}[t]
\renewcommand{\arraystretch}{1.25} 
\centering
\caption{\label{tab:near_field_setup} Simulated environmental setup for the near-field acoustic field prediction application.}
\begin{tabular}{ll}
\toprule \toprule
{Parameters} & {Value}\\
\midrule
Environmental model & 3D \\
Frequency & 5 kHz \\
Seabed & Sandy clay\\
Bathymetry & Range-independent\\
Water depth & 30 m\\
Source depth & 15 m\\
Sound speed & 1,541 m/s \\
Distance between source and AOI & 100 m \\
Dimension of AOI & \update{50~m~$\times$~28~m}\\
Number of training data & 116 \\
Number of validation data& 51 \\
Number of test data \update{in AOI} & 561,561 \\
Number of rays in RBNN & 60\\
\bottomrule
\end{tabular}
\end{table}

The setup of the simulated environment is shown in Fig.~\ref{fig:sph_schemetic} and summarized in Table~\ref{tab:near_field_setup}. A profiling float performs 2 profiles across a 50~m $\times$ 28~m AOI at a distance of 100~m from a 5~kHz source deployed at a depth of 15~m to collect 167 acoustic field measurements. \update{We use the Pekeris ray model~\cite{mandar_pekeris} to generate synthetic data.} We use 70\% of the collected measurements to train the RBNN, and aim to estimate the acoustic field in the entire AOI. We benchmark the field estimation performance of the RBNN against GPR and DNN with the same model configurations as discussed in Section~\ref{sec:far_field_est}. 

A dense test dataset of 561,561 data points on a 0.05~m spacing grid covering the AOI is generated to evaluate the field prediction performance. As RBNN and DNN may be sensitive to random initialization, we carry out \update{20} Monte Carlo simulations for each, and present the results with the best validation error. The \update{initial values of} hyper-parameters in the GPR kernel are tuned to yield the best validation error. \update{We then use the L-BFGS technique to further optimize hyper-parameters by minimizing the training error of the GPR model.} To evaluate model robustness, we \update{add} random measurement errors in the source location (0.3~m in horizontal directions, 0.1~m in depth), measurement locations (maximum of 0.4~m in all directions) and water depth (1~m) of AOI.

\begin{table}[t]
\renewcommand{\arraystretch}{1.25} 
\caption{\label{tab:sph_error} RMS test error for the near-field acoustic field prediction application.}
\centering
\
\begin{tabular}{lll}
\toprule\toprule
{Method} & \multicolumn{2}{c}{{RMS test error} (dB)} \\
\cline{2-3}
  & {Error-free data} & {Noisy data}\\
\midrule
RBNN & \update{1.889}  & \update{6.219} \\
GPR & \update{7.119} & \update{7.066} \\
DNN & \update{7.074}  & \update{7.127}\\
\bottomrule
\end{tabular}
\end{table}

The RMS test errors of the estimated fields within the AOI by the three models are shown in Table~\ref{tab:sph_error}. The acoustic field patterns within AOI estimated by the three approaches are shown in Fig.~\ref{fig:sph_res} (c), (d) and (e). Fig.~\ref{fig:sph_res} (f), (g) and (h) show extrapolated fields by the three models. The RBNN model is able to predict and extrapolate the spatially fast-varying field patterns well, even with a much smaller training dataset as compared to the far-field acoustic field prediction application in the previous section. However, the GPR and DNN show poor performance in terms of the estimated field pattern and RMS test error, for both interpolation and extrapolation of the acoustic field. The results highlight the data-efficiency of the RBNN model -- the model can effectively incorporate knowledge of channel geometry and therefore train with very little data. The conventional GPR and DNN models, on the other hand, fail to predict field patterns as they do not benefit from partial environmental knowledge.

As one would expect, measurement noise worsens the prediction accuracy for the RBNN model. The sensitivity of field estimation to position errors is summarized in Table~\ref{tab:sph_pos_error}. However, the qualitative field pattern can still be recovered even with large measurement errors as seen in Fig.~\ref{fig:sph_res} (i).

\begin{table}[t]
\centering
\caption{\label{tab:sph_pos_error} Sensitivity of RMS test error of field estimation to random position error for the near-field acoustic field prediction application.}
\begin{threeparttable}
\renewcommand{\arraystretch}{1.25} 
\begin{tabular}{ll}
\toprule \toprule
 {Maximum position error\tnote{1}}~(m)  & {RMS test error}~(dB)\\
\midrule
0.0 &  \update{1.889}\\
0.1$\sqrt{3}$ & \update{3.401} \\
0.2$\sqrt{3}$ & \update{4.665} \\
0.3$\sqrt{3}$ & \update{5.423} \\
0.4$\sqrt{3}$ & \update{6.219} \\
\bottomrule
\end{tabular}
\begin{footnotesize}
\begin{tablenotes}
  \item[1] Maximum position error per dimension $\times \sqrt{3}$.
  \end{tablenotes}
\end{footnotesize}
\end{threeparttable}
\end{table}

\subsection{Geo-acoustic inversion for a seabed reflection model}\label{sec:rc_curve}

\begin{figure*}[t]
\includegraphics[width=\linewidth]{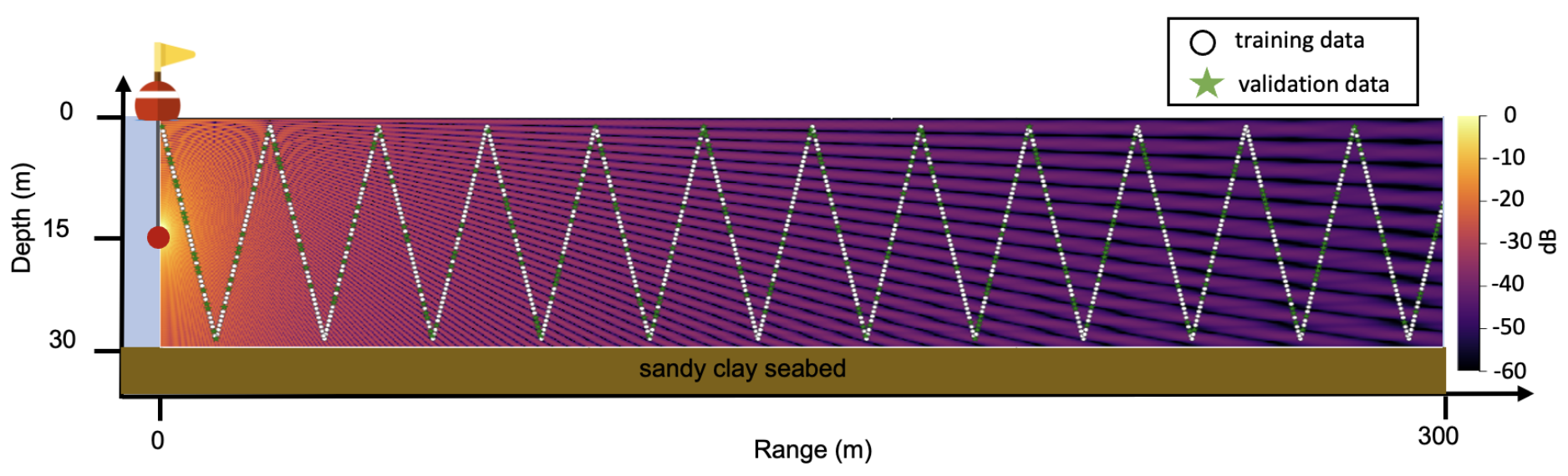}
\caption{\update{Simulated environment for geo-acoustic inversion for seabed reflection model. The trajectory of the profiling float can be seen in terms of the training data points. The ground truth field pattern within the AOI is also shown.}}
\label{fig:rc_schemetic} 
\end{figure*}

The third application we demonstrate is to extract a seabed reflection model from acoustic measurements. \update{While our proposed model primarily targets forward problems, it also has the potential to solve geo-acoustic inversion problems as a by-product.}

Seabed reflection depends on the seabed structure and material properties, which are often unknown. In applications where multipath arrivals overlap and cannot be separated, we cannot measure the reflection coefficient directly. Our proposed recipe can, however, learn a reflection model from observed total transmission loss at a number of observation points. We use the RBNN model from Section~\ref{sec:sph_with_geo} together with the RCNN layer, where the RCNN models the unknown reflection coefficient (as a function of reflection angle). By training the composite spherical wave model described in \eqref{eq:sph_amplitude}, we can recover a trained RCNN as a model for the seabed reflection. 

We assume that the channel geometry and source location are known. A profiling float is employed to perform \update{12} profiles through a 300~m $\times$ \update{28}~m AOI from a 5~kHz source deployed at a depth of 15~m. The simulation setup is illustrated in Fig.~\ref{fig:rc_schemetic}. A total of \update{1,112} acoustic field measurements are collected along the trajectory, and 70\% of them are used to train the composite RBNN model. The environmental setup is similar to the near-field acoustic field prediction application in Section~\ref{sec:sph_field_est}, and the synthetic data \update{are} generated using the \update{Pekeris ray} model. We pre-calculate the nominal arrival directions and propagation distances, and use acoustic data to optimize the trainable model parameters~$\mathcal{T}$:
\begin{equation}
\update{\mathcal{T} \equiv \left ( \boldsymbol{R} \right )}.
\end{equation}
\update{We use \eqref{eq: spherical_loss_func} as the loss function to learn the reflection model.}

{\small
\begin{table*}[t]
\renewcommand{\arraystretch}{1.2} 
\normalsize
\centering
\caption{\label{tab:infer_rc}\update{Effect of measurement position error on the seabed reflection model}.}
\begin{threeparttable}
\centering
\begin{tabular}{>{\centering\arraybackslash} p{4cm} >{\centering\arraybackslash} m{6cm} >{\centering\arraybackslash} m{6cm}}
\toprule \toprule
{Maximum position error}~(m)& {Reflection coefficient} & {Reflection phase shift}\\
\midrule
0.00 &  {\includegraphics[width=0.31\textwidth]{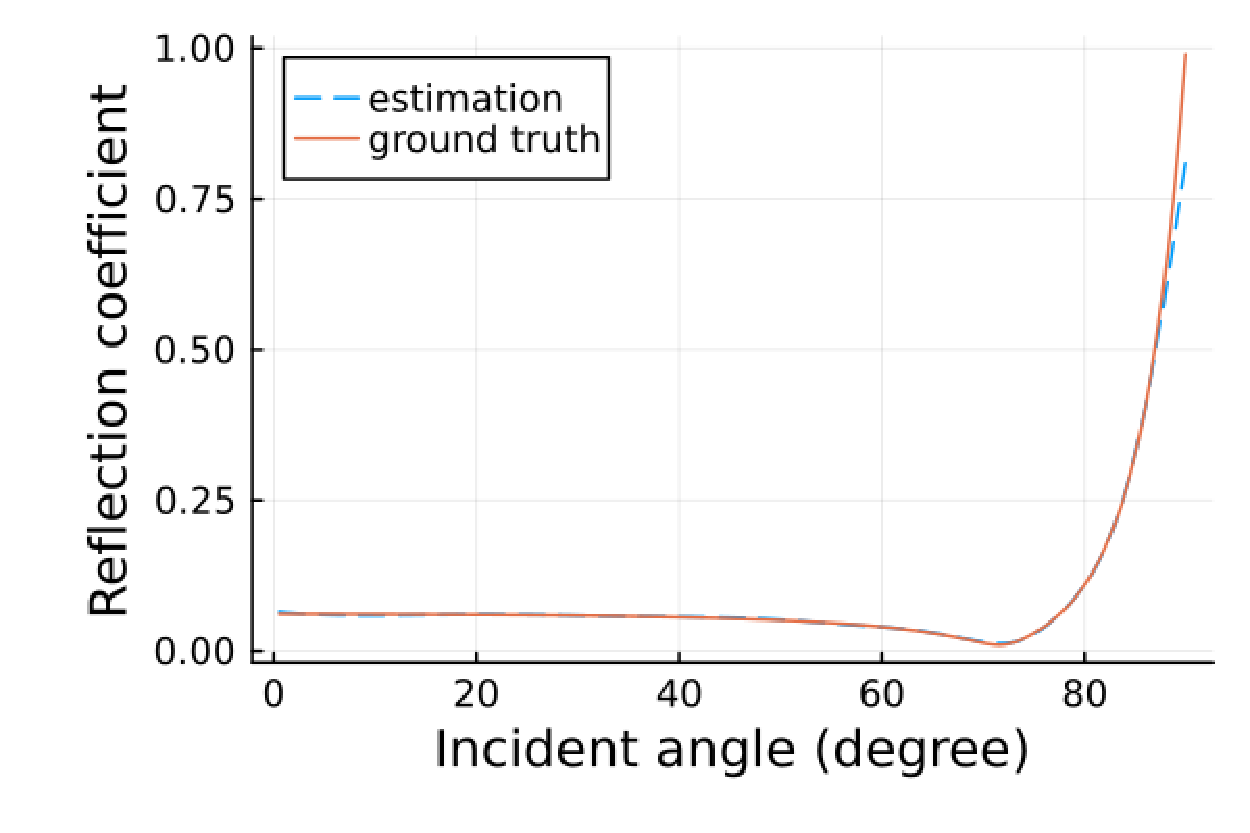}}&  {\includegraphics[width=0.31\textwidth]{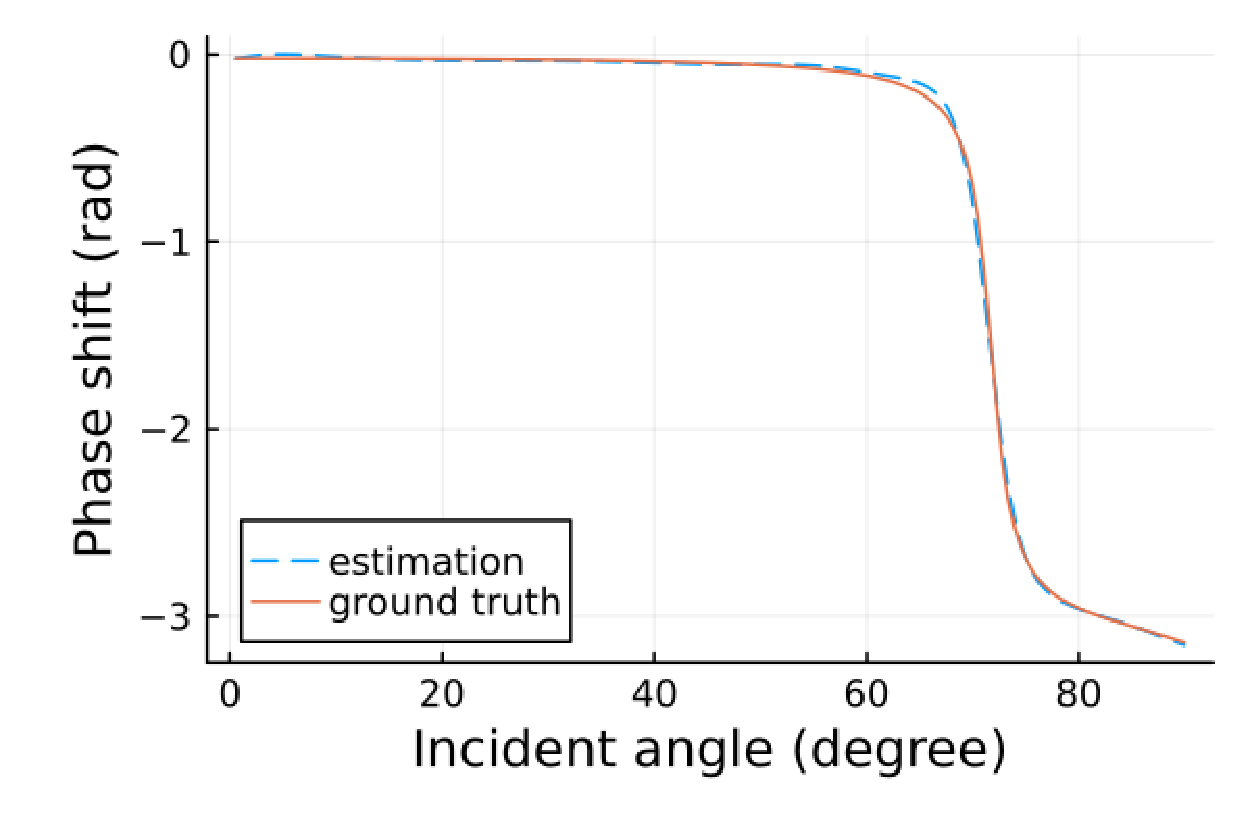}}\\
\hline
0.01$\sqrt{3}$ & {\includegraphics[width=0.31\textwidth]{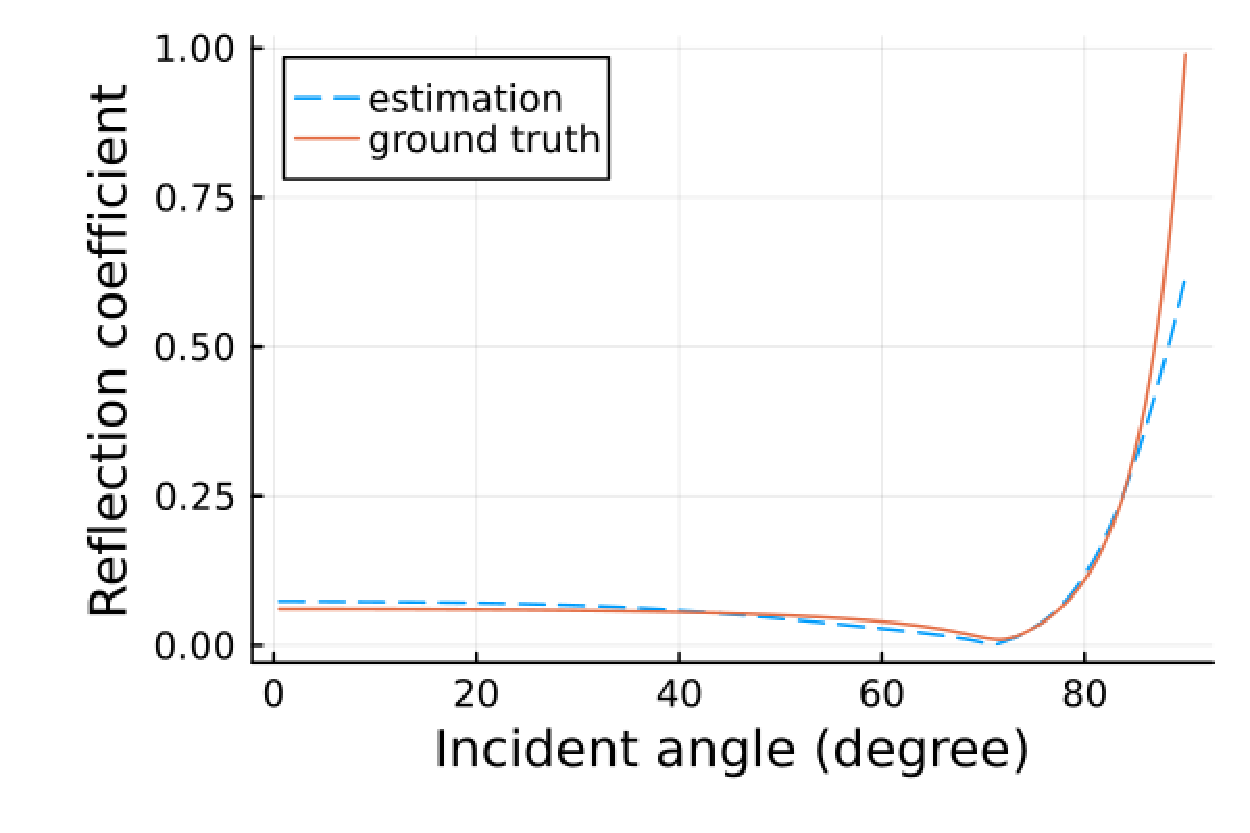}}&  {\includegraphics[width=0.31\textwidth]{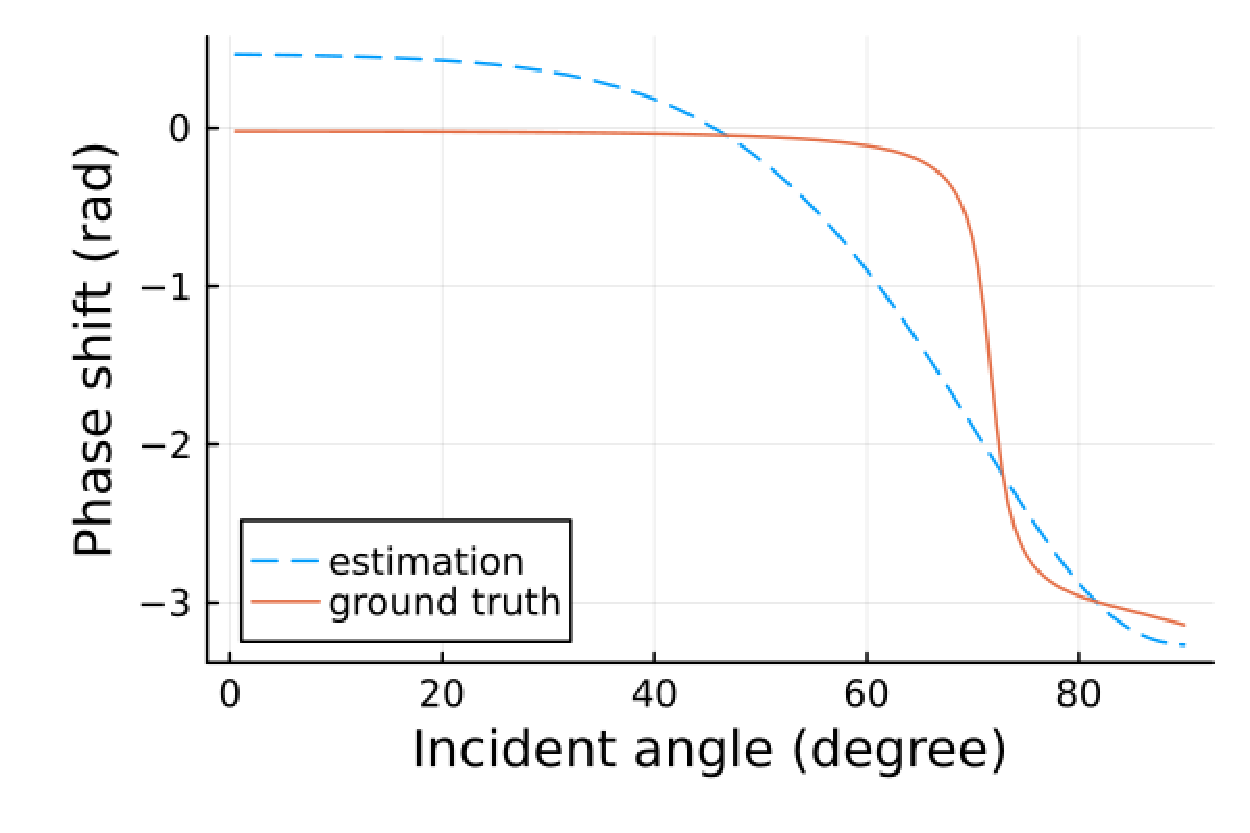}}\\
\hline
0.05$\sqrt{3}$ & {\includegraphics[width=0.31\textwidth]{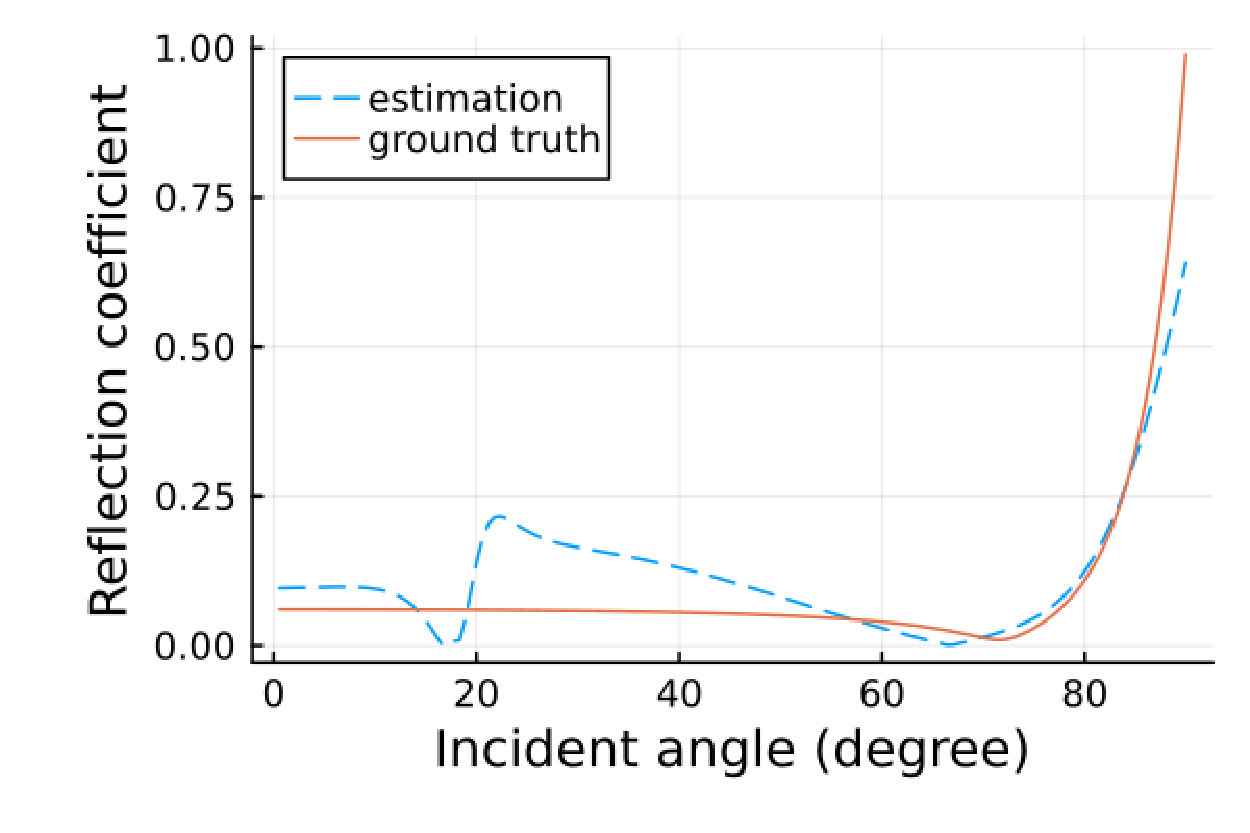}}&{ \includegraphics[width=0.31\textwidth]{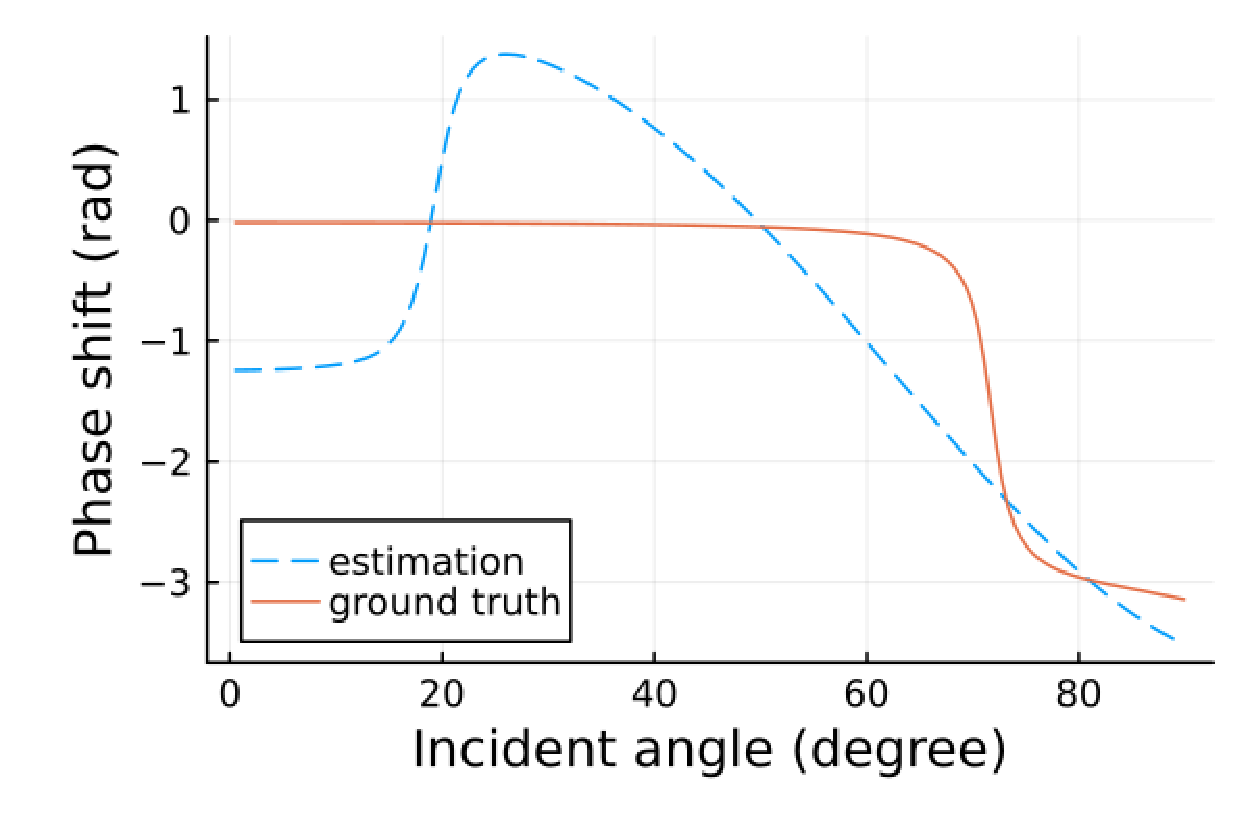}}\\
\hline
0.10$\sqrt{3}$ & {\includegraphics[width=0.31\textwidth]{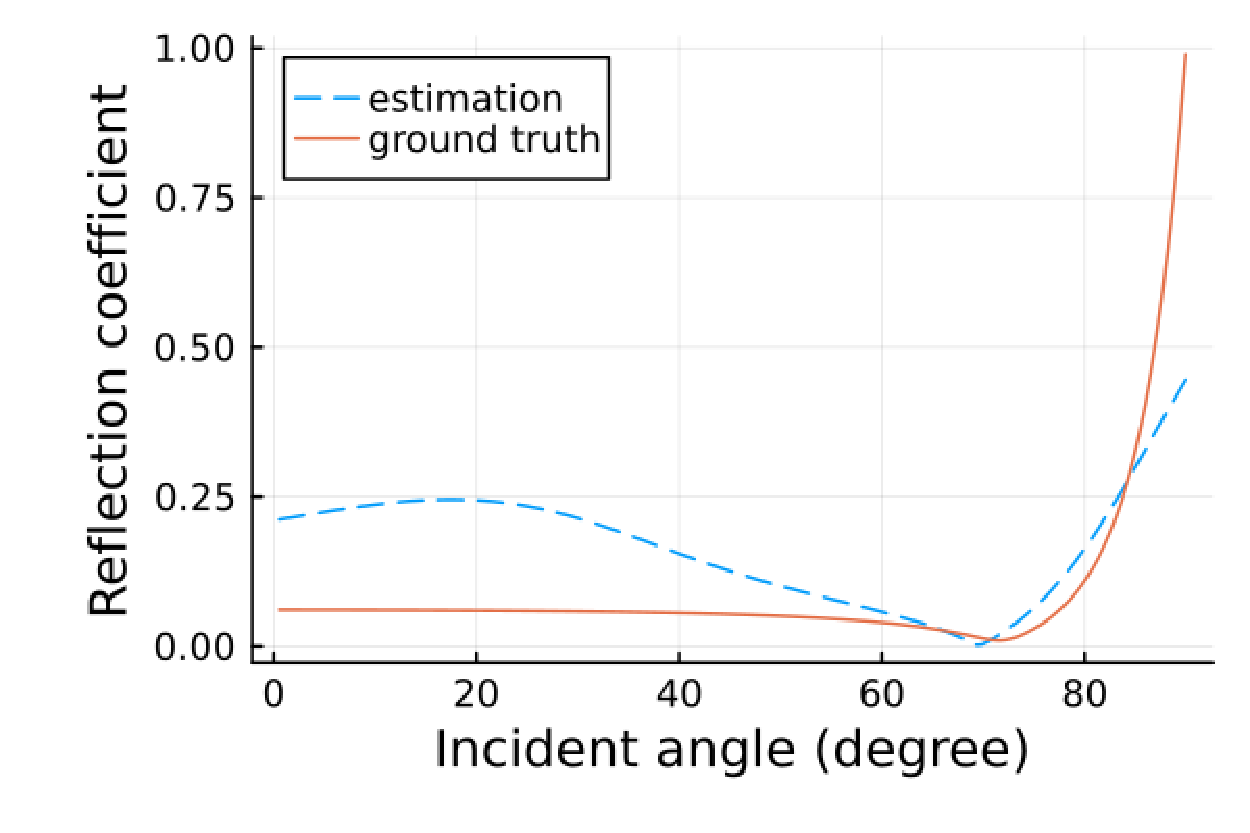}}& {\includegraphics[width=0.31\textwidth]{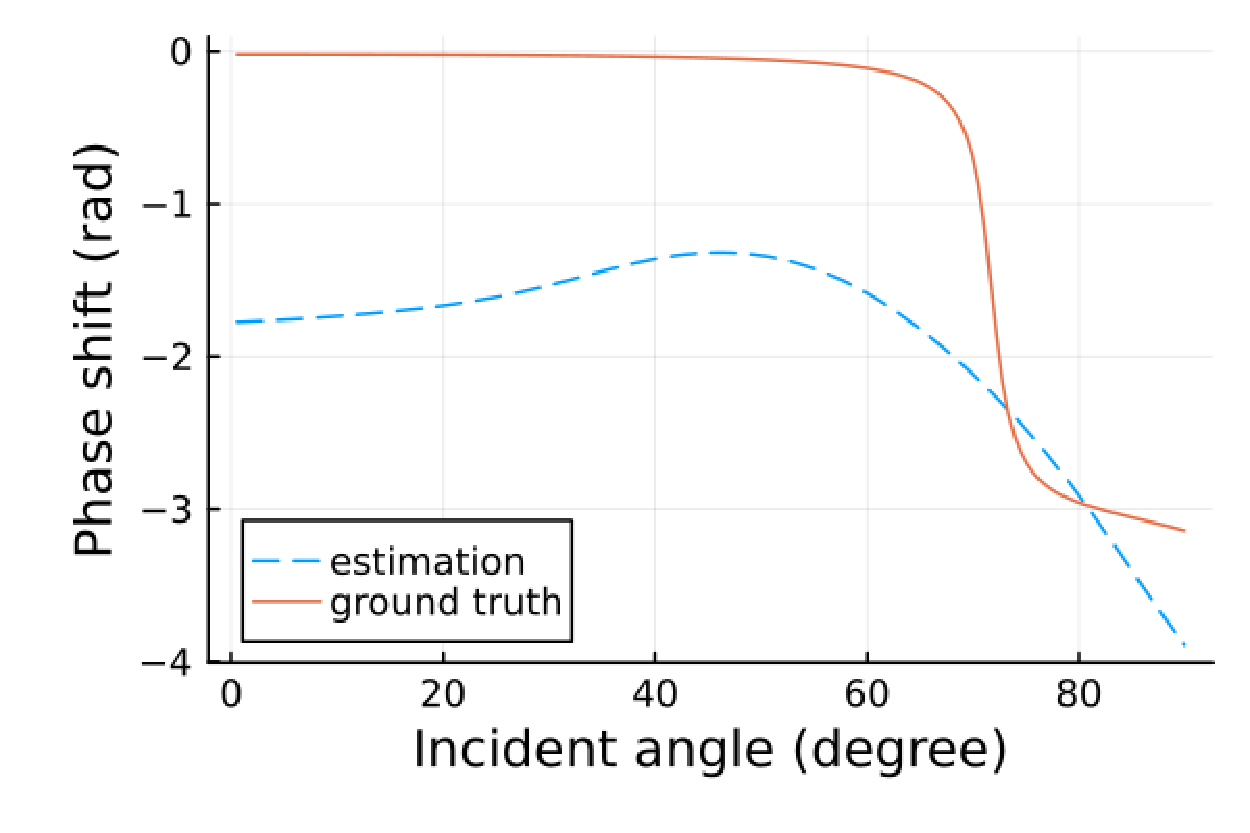}}\\
\bottomrule
\end{tabular}
\end{threeparttable}
\end{table*}
}

In Table~\ref{tab:infer_rc}, we present the inferred reflection coefficient curves and phase shift curves for various \update{amounts} of position measurement error. In an ideal scenario with no measurement errors, we can accurately recover the seabed reflection model. The modeling errors increase with the amount of position measurement error, \update{with phase of the reflection coefficient being more sensitive to errors than its amplitude}. The effect of measurement error can be partially mitigated by increasing the size of the training dataset.

\subsection{Geo-acoustic inversion for seabed properties}

The last application we shall consider is a geo-acoustic inversion problem, where we wish to determine geo-acoustic seabed properties from acoustical field measurements. 
We consider a simple Rayleigh reflection model to illustrate the idea. The complex Rayleigh reflection coefficient is given by~\cite{Brekhovskikh2003}:
\begin{equation}
\Gamma = \frac{\rho_\text{r}\cos{\gamma} - \sqrt{(\frac{\bar{\delta}}{c_\text{r}})^2 - \sin{\gamma}^2}}{\rho_\text{r}\cos{\gamma} + \sqrt{(\frac{\bar{\delta}}{c_\text{r}})^2 - \sin{\gamma}^2}},
\label{eq:reflection_coeff}
\end{equation}
where 
\begin{subequations}
\begin{align}
\bar{\delta} = 1 + i\delta,\\
\rho_\text{r} = \frac{\rho_\text{seabed}}{\rho_\text{seawater}},\\
c_\text{r} = \frac{c_\text{seabed}}{c_\text{seawater}},
\end{align}
\end{subequations}
where $\delta$ denotes dimensionless seabed absorption coefficient, $\rho_\text{r}$ denotes relative density, $c_\text{r} $ represents relative sound speed. We assume $\rho_\text{r}$, $c_\text{r}$ and $\delta$ are unknown and to be determined from acoustic field measurements.

\begin{figure}[t]
\centering
\includegraphics[width=0.75\linewidth]{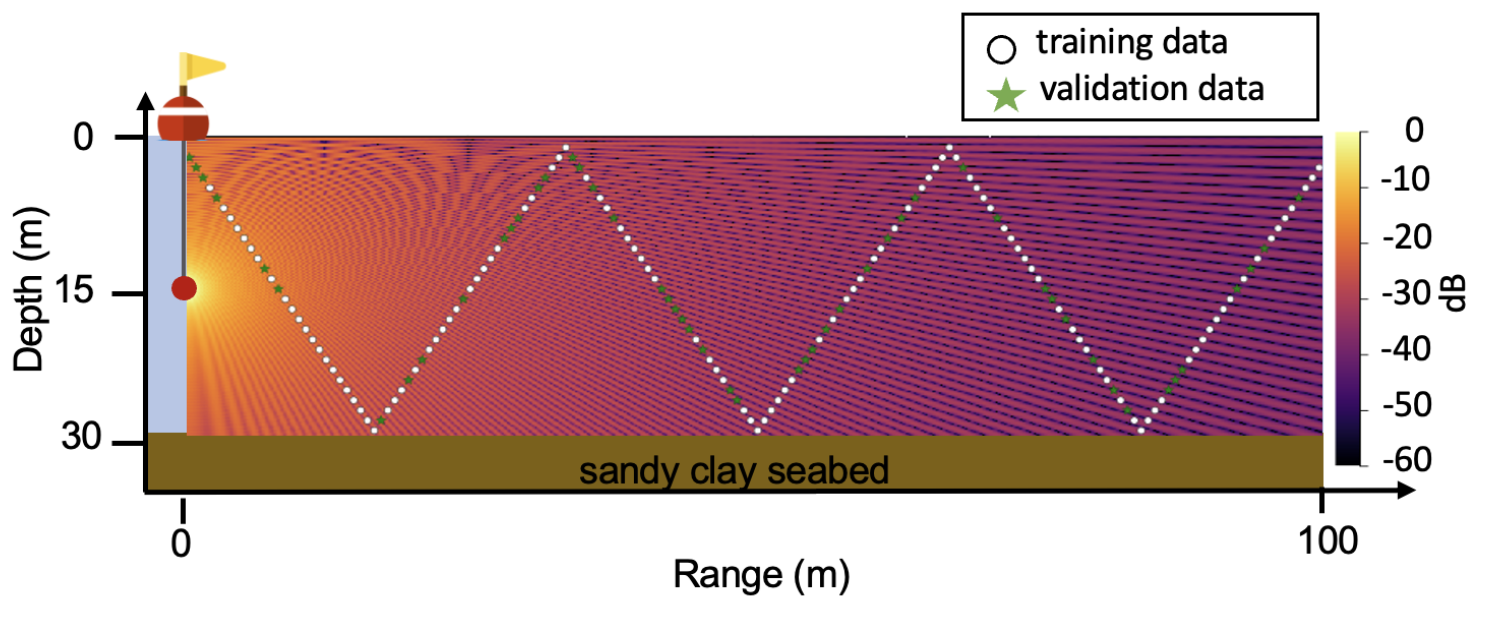}
\caption{\update{Simulated environment for the geo-acoustic inversion of seabed properties. The trajectory of the profiling float can be seen in terms of the training data points. The ground truth field pattern within the AOI is also shown.}}
\label{fig:para_schemetic} 
\end{figure}

We assume that the source and receiver locations, as well as the channel geometry are known. An \update{profiling} float is employed to take $166$ acoustic measurements over 100~m $\times$ \update{28}~m AOI along a zig-zag trajectory from a 5~kHz acoustic source deployed at a depth of 15~m. As in previous applications, 70\% of the measurements are used to train the RBNN model, while the balance 30\% is used for validation. Fig.~\ref{fig:para_schemetic} depicts the simulated environment and the sampling trajectory of the \update{profiling} float.

In Section~\ref{sec:rc_curve}, we modeled the angle-dependent complex reflection coefficient using a RCNN. While this is useful for acoustic propagation modeling, this approach does not yield estimates of geo-acoustic properties such as $\rho_\text{r}$, $c_\text{r}$ and $\delta$. We therefore replace the RCNN layer in~\eqref{eq:sph_amplitude} with the expression for complex reflection coefficient from~\eqref{eq:reflection_coeff}, and train the resultant composite RBNN.

\update{We use \eqref{eq:p_loss_func} as the loss function to learn the best-fitted geoacoustic parameters.} The set of trainable parameters for this RBNN is:
\begin{equation}
\update{\mathcal{T} \equiv \left (\rho_\text{r}, c_\text{r}, \delta \right)}.
\end{equation}

Table~\ref{tab:tank_para_error} summarizes the estimated values and percentage error for the three unknown geo-acoustic parameters for various levels of position measurement errors. With accurate measurements, the model is effective in accurately determining the geo-acoustic parameters. In the presence of position errors, we need to increase the training data size to improve model robustness. The robustness of geo-acoustic inversion depends on the sensitivity of the acoustic field to each geo-acoustic parameter. In this example, $\rho_\text{r}$ and $c_\text{r}$ affect the acoustic field more strongly than $\delta$. Increasing training dataset size improves the robustness of the estimates of $\rho_\text{r}$ and $c_\text{r}$, but much less so for $\delta$.

\begin{table}[t]
\renewcommand{\arraystretch}{1.25} 
\caption{\label{tab:tank_para_error} Estimated seabed parameters as a function of maximum position measurement error.}
\centering
\begin{threeparttable} 
\begin{tabular}{llllll}
\toprule \toprule
Maximum & Training & \update{Distance} & {$\rho_\text{r}$} & {$c_\text{r}$} & {$\log(\delta)$}~(dB) \\
 error~(m) & \update{data size} & \update{travelled (m)} & [{\% error}] & [{\% error}] & [{\% error}]\\
\midrule
0.00  & \update{167} & \update{100} & 1.147 &0.985 & -2.616\\
 &  & &[0.0\%] & [0.0\%] & [0.0\%]\\
\hline
0.01$\sqrt{3}$ & \update{167} & \update{100} & \update{1.125} & \update{0.989} & \update{-2.444}\\
 &  & &[\update{-1.9\%}] & [\update{0.4\%}] &[\update{6.6\%}]\\
\hline
0.05$\sqrt{3}$&  \update{250} & \update{150}& \update{1.036} & \update{0.998} & \update{-2.728}\\
& & & [\update{-9.7\%}] & [\update{1.3\%}] & [\update{-4.3\%}]\\
\hline
0.10$\sqrt{3}$ & \update{334} & \update{200} & \update{1.209} & \update{0.962} & \update{-1.485}\\
&& &[\update{5.4\%}] & [\update{-2.3\%}] & [\update{43.2\%}]\\
\hline
0.20$\sqrt{3}$  & \update{375} & \update{250} & \update{1.084} & \update{0.993} & \update{-2.742}\\
&&&[\update{-5.5\%}]& [\update{0.8\%}] & [\update{-4.8\%}]\\
\hline
0.50$\sqrt{3}$ & \update{417} & \update{300} & \update{1.249} & \update{0.962} & \update{-1.432}\\
&& &[\update{8.9\%}] & [\update{-2.3\%}] & [\update{45.3\%}]\\
\bottomrule
\end{tabular}
\end{threeparttable}
\end{table}

\section{Experimental Validation}\label{sec:experiment}

To further validate the acoustic field estimation performance of the proposed framework, we \update{undertook} a controlled experiment in a water tank. This allows us to make careful repeatable measurements to validate the method -- something that is very difficult to do at sea due to time variability.

The tank environment is strongly reverberant and surprisingly complicated to model. While acoustic rays in the rectangular geometry can be modeled with a 3D ray-tracer, multiple reflections lead to strong sensitivity to minor geometrical irregularities of the tank wall. The tank walls are made of an inhomogeneous composite material (fiberglass) with complicated reflection properties. This provides us \update{with} a challenging acoustic propagation modeling problem to demonstrate our proposed method.

Before undertaking experimental validation, we developed a simplified simulation model of the tank to establish the feasibility of applying our method to the tank environment. The simulation results are presented in Section~\ref{sec:simulated_feasibility_study}. Once we had established the feasibility and developed an understanding of what performance we might expect, we undertook experimental validation in the tank. The results from the experiment are presented in Section~\ref{sec:controlled-experiment}.

\subsection{Feasibility study}\label{sec:simulated_feasibility_study}

\begin{figure}[t]
\begin{subfigure}[t]{0.45\textwidth}
\includegraphics[width=\linewidth]{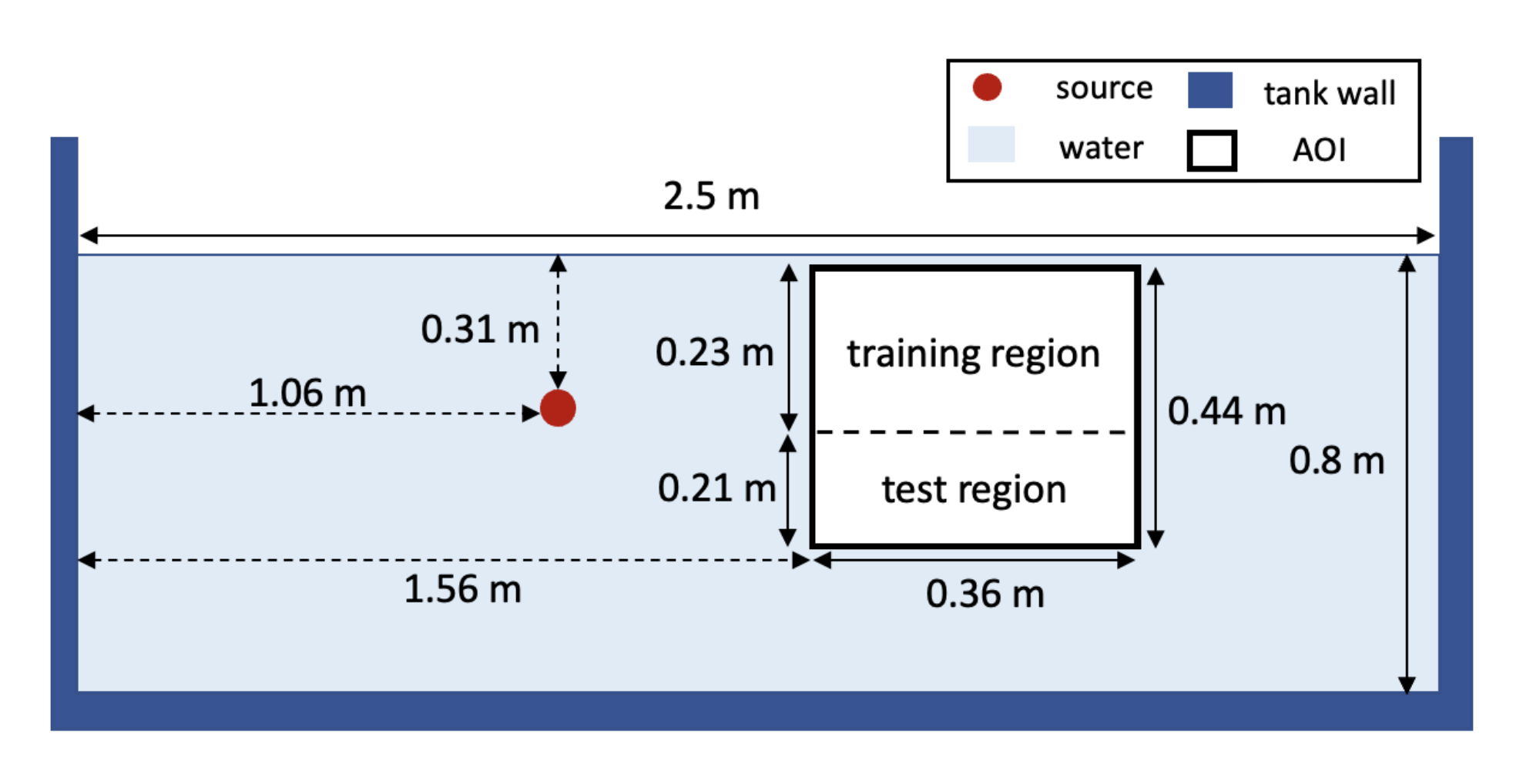}
\caption{Side view} 
%\vspace{1\baselineskip}
\end{subfigure}
\begin{subfigure}[t]{0.45\textwidth}
\includegraphics[width=\linewidth]{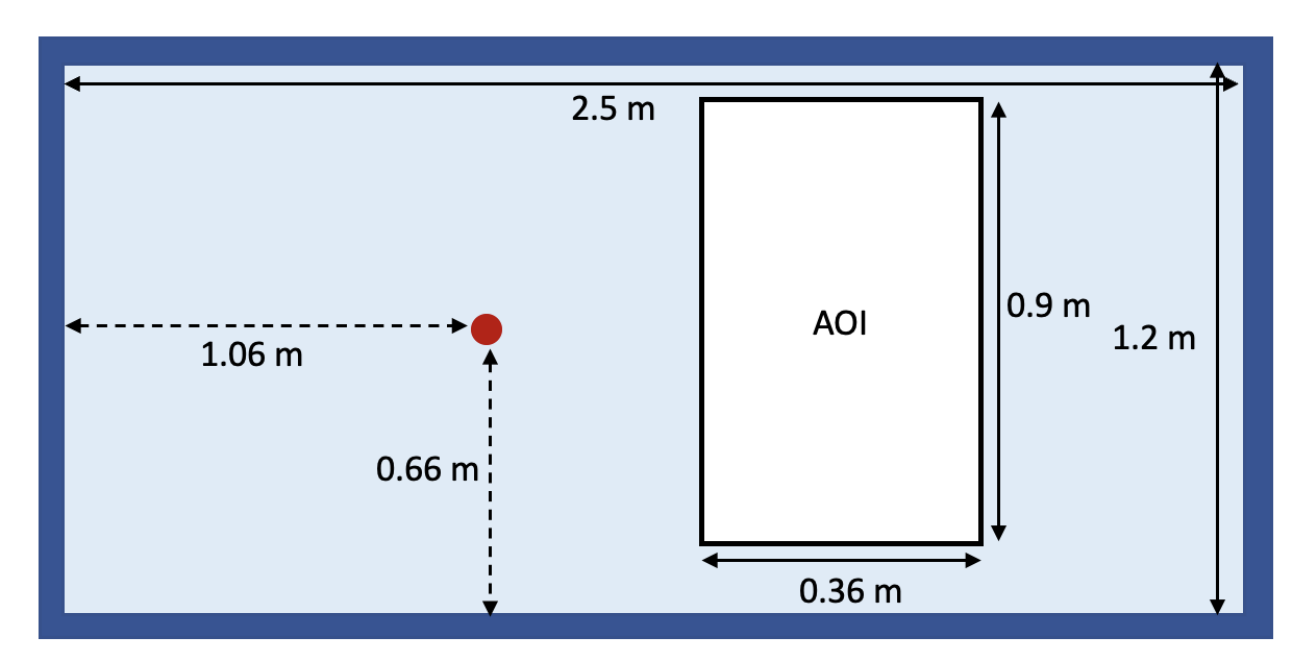}
\caption{Top view} 
\end{subfigure}
\caption{Tank experiment setup.}
\label{fig:tank}
\end{figure}

We simulate a 3D water tank environment with the dimension of $2.5$~m~$\times$~$1.2$~m~$\times$ $0.8$~m and a $10$~kHz CW signal source, as illustrated in Fig.~\ref{fig:tank}. A 0.36~m~$\times$~$0.9$~m~$\times$ $0.44$~m AOI is located $0.5$~m from \update{the} source. The sound speed is assumed to be $1,505$~m/s, in accordance with conductivity and temperature measurements in our tank. We split the AOI into \update{non-overlapping training and test regions}. The training and validation data (250 and 28 data points respectively) are obtained from the training region, whereas the test region is used to test (222 data points) how well the model extrapolates beyond the training region. We adopt the spherical wave formulation with the knowledge of geometry based on~\eqref{eq:sph_amplitude} and~\eqref{eq:sph_nominal_set}\footnote{\update{We assume $k$ is known as sound speed $c$ is measurable in this experiment.}} to predict the field in AOI. The loss function we minimize is~\eqref{eq: spherical_loss_func}.

We adopt a geometrical ray model to simulate the acoustic propagation in the tank environment, and generate synthetic acoustic measurements\footnote{The acoustic measurements are shown in Volts, as we measure the pre-amplified output from the hydrophones in Volts during the experiment. These can be converted to $\mu$Pa by multiplying by the gain-corrected acoustic sensitivity of the hydrophone.} within the AOI. We assume the water-air interface to be a perfect reflector with a reflection coefficient of $-1$. We adopt a simple tank sidewall and bottom reflection model, and assume the reflection coefficient to be given by~\eqref{eq:reflection_coeff}, with $\rho_\text{r} = 1.5$, $c_\text{r} = 0.9$ and $\delta = 0.0$. For benchmarking, we use GPR and DNN similar to those described in Section~\ref{sec:far_field_est}. We generate a dense test dataset of $30,303$ points over the entire AOI with a resolution of $0.01$~m in range and width, and $0.05$~m in depth. It is not practical to collect such a dense dataset during the \update{latter} experiment, and so we also generate a sparse test dataset of $222$ points in the test region for later benchmarking of the experimental results.
\begin{figure*}
\begin{subfigure}[t]{\textwidth}
\includegraphics[width=\linewidth]{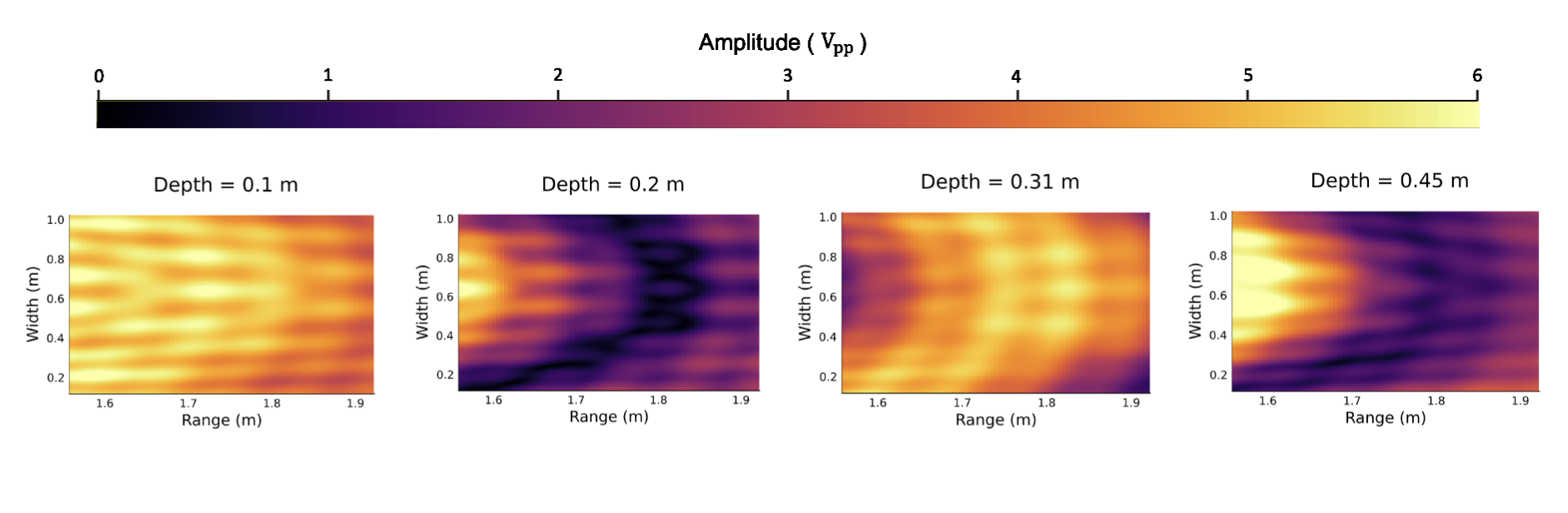}
\caption{Ground truth field within the AOI}
\vspace{3ex}
\end{subfigure}
\begin{subfigure}[t]{\textwidth}
\includegraphics[width=\linewidth]{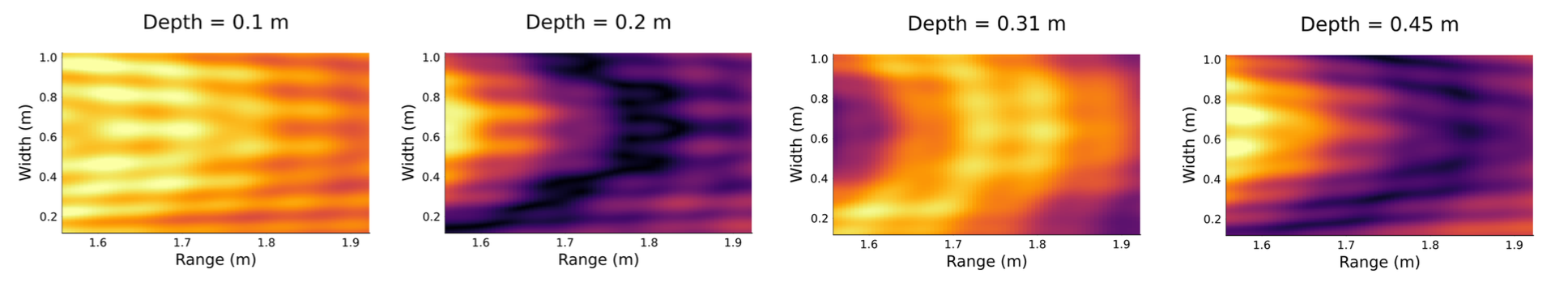
}
\caption{\update{RBNN field estimates within AOI}}
\vspace{3ex}
\end{subfigure}
\begin{subfigure}[t]{\textwidth}
\includegraphics[width=\linewidth]{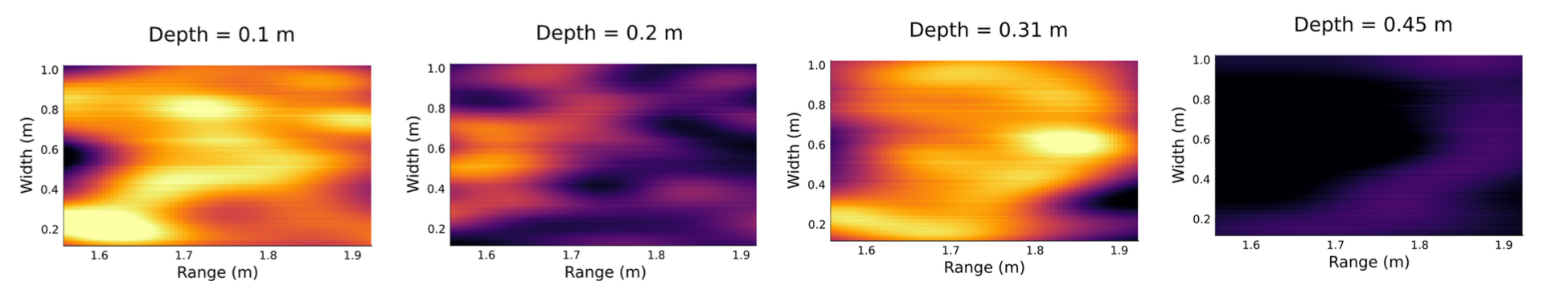}
\caption{\update{GPR field estimates within AOI}}
\vspace{3ex}
\end{subfigure}
\begin{subfigure}[t]{\textwidth}
\includegraphics[width=\linewidth]{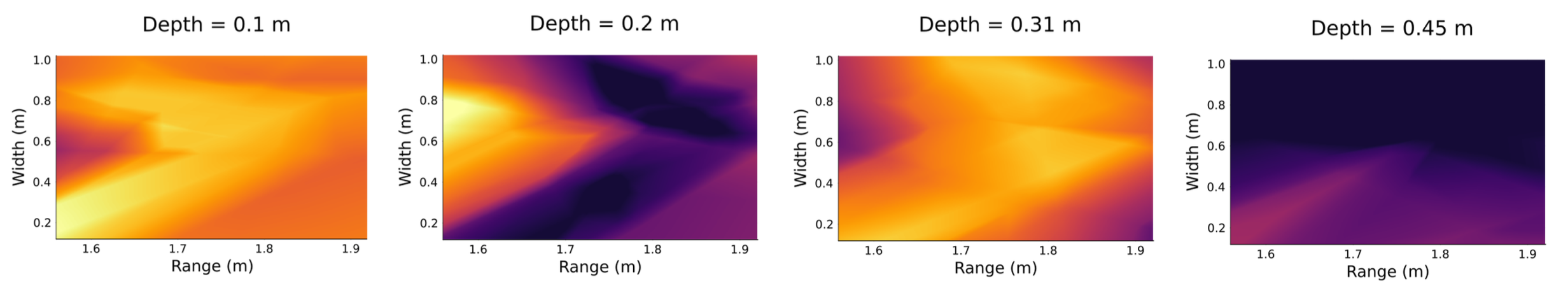}
\caption{\update{DNN field estimates within AOI}}
\vspace{3ex}
\end{subfigure}
\caption{Ground truth and estimated acoustic field at four different depths. The depth of $0.45$~m is in the test region, where no training data \update{were} made available to the three models. The other three depths are in the training region.}
\label{fig:tank_sim_local}
\end{figure*}

\begin{figure*}
\begin{subfigure}[t]{1.035\textwidth}
\includegraphics[width=\linewidth]{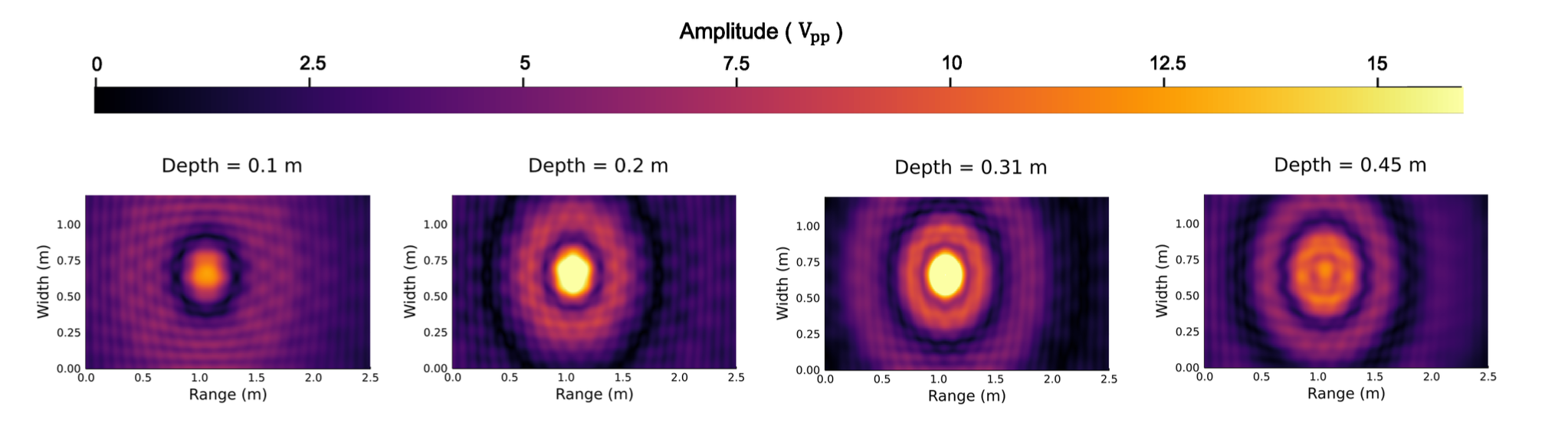}
\caption{Ground truth field in \update{the AOI and} extended region}
\vspace{3ex}
\end{subfigure}
\begin{subfigure}[t]{\textwidth}
\includegraphics[width=\linewidth]{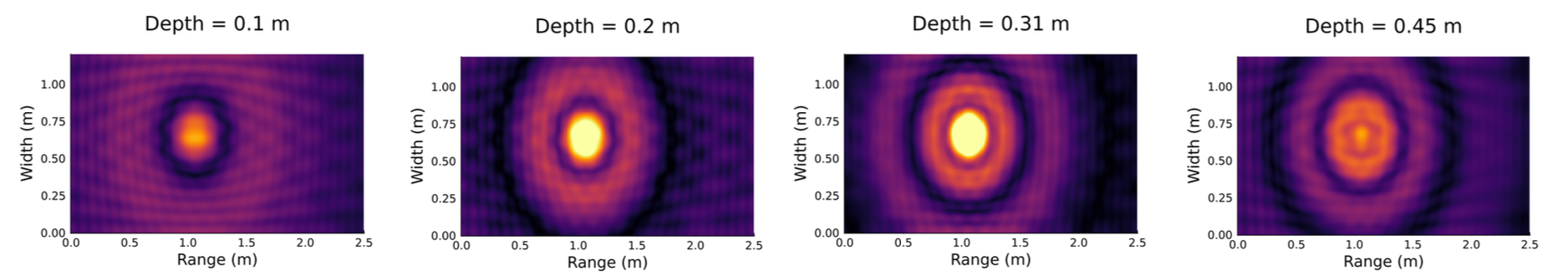}
\caption{\update{RBNN extrapolated field estimates}}
\vspace{3ex}
\end{subfigure}
\begin{subfigure}[t]{\textwidth}
\includegraphics[width=1.01\linewidth]{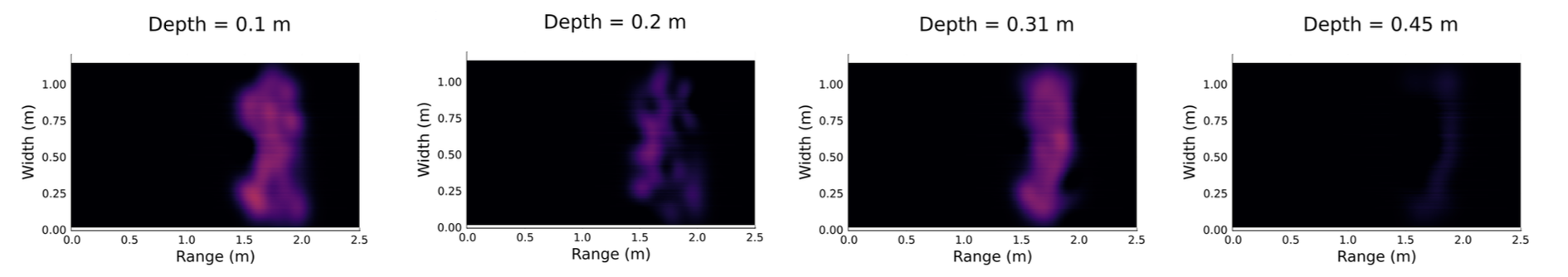}
\caption{\update{GPR extrapolated field estimates}}
\vspace{3ex}
\end{subfigure}
\begin{subfigure}[t]{\textwidth}
\includegraphics[width=\linewidth]{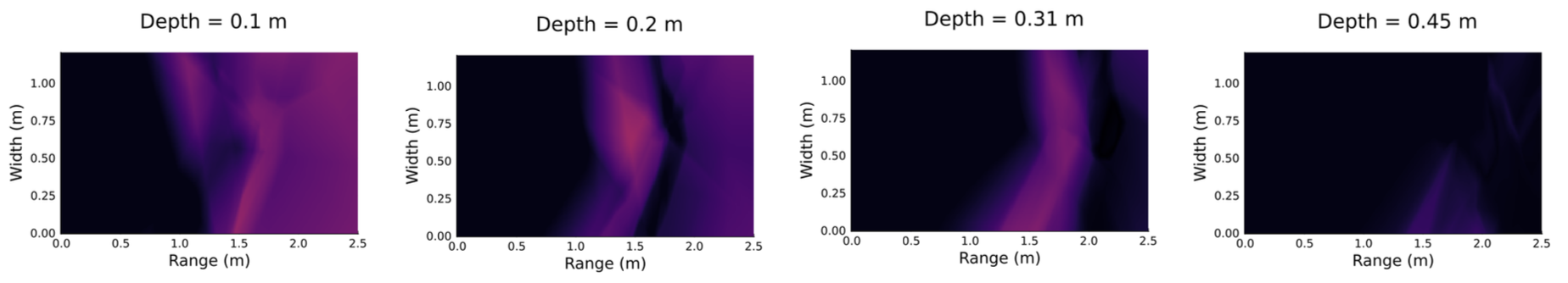}
\caption{\update{DNN extrapolated field estimates}}
\vspace{3ex}
\end{subfigure}
\caption{Ground truth and extrapolated field within the entire tank at four different depths.}
\label{fig:tank_sim_full}
\end{figure*}

Since the measurement accuracy of tank dimensions and transducer locations in the tank is limited, we introduce measurement errors in the tank size, source location and measurement locations in simulation too. The simulated tank dimensions are mismatched from the geometrical knowledge available to our algorithm by $0.010$~m, $0.015$~m and $0.020$~m in the three dimensions. The source location deviates by $0.02$~m from the location provided to the algorithm. Due to practical considerations, the measurement errors in shallower hydrophone locations in our tank \update{are} expected to be less than that for deeper locations. We therefore introduce a random error of up to $0.02$~m per dimension for acoustic measurements with depths shallower than $0.36$~m, and $0.04$~m per dimension for deeper locations. We calculate the nominal incoming ray directions and propagation distances prior to the training. We allow our RBNN model to train the error to the nominal directions and propagation distances to cope with the erroneous source location and tank size measurements, as discussed in Section~\ref{sec:sph_with_geo}. To allow for a few measurement outliers during the experiment, we opt to minimize the mean absolute error in the training process, rather than the RMS error. This encourages the model to focus on fitting the majority of the training data well, and ignore a few outliers.

\begin{table}[t]
\renewcommand{\arraystretch}{1.25} 
\centering
\caption{\label{tab:tank_sim_err} MATE of the estimated acoustic field for the feasibility study.}
\begin{tabularx}{0.5\columnwidth}{X X X}
\toprule \toprule
\multirow{2}{*}{{Method}} &  \multicolumn{2}{c}{{MATE}~(V$_\text{pp}$)}\\
\cline{2-3} & Sparse & Dense \\
\midrule
RBNN & \update{0.351} & \update{0.292} \\
GPR  & \update{1.846} & \update{1.196} \\
DNN  & \update{1.426} & \update{0.919}  \\
\bottomrule
\end{tabularx}
\end{table}

The rich multipath in the simulated water tank environment yields a complicated field pattern. Cross-sections of the ground truth field and the estimated field at four different depths within the AOI are shown in Fig.~\ref{fig:tank_sim_local}. Note that the estimated field at the depth of $0.45$~m is extrapolated as none of the training data or validation data falls in this test region. We see that the RBNN model can recover and extrapolate the field reasonably well, whereas the GPR and RBNN methods fail to do so. The mean absolute test error~(MATE) of the sparse and dense test datasets is shown in Table~\ref{tab:tank_sim_err}. 
The sparse test error and dense test error are based on error-free measurements. The two types of test errors are in \update{a} similar range for all of the three models. This suggests that the sparse test error is a representable measure of field estimation performance. We also extrapolate the field to the entire water tank environment as shown in Fig.~\ref{fig:tank_sim_full}. Not surprisingly, the \update{two} classical data-driven techniques \update{-- GPR and DNN} fail to extrapolate the field in the region away from the AOI, whereas the RBNN model can generalize well and predict the field in the entire water tank.

\begin{figure}[t]
\begin{center}
\includegraphics[width=0.5\textwidth]{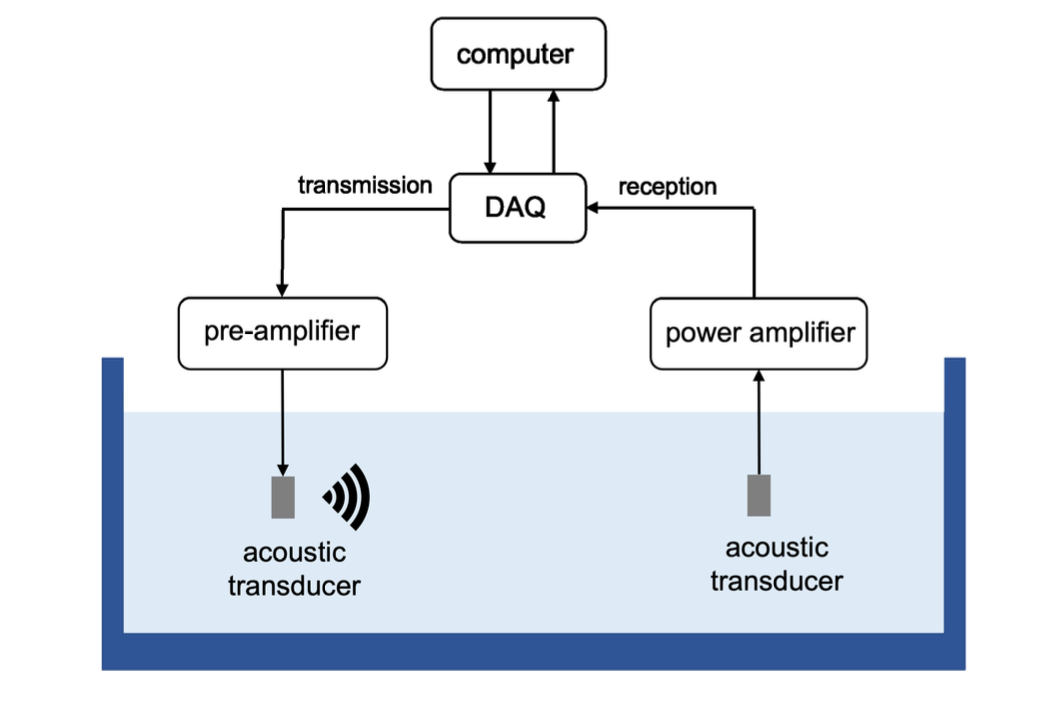}
\end{center}
\caption{Equipment setup for the controlled experiment. \label{fig:tank_flow}}
\end{figure}

\begin{figure}[t]
\begin{center}
\includegraphics[width=0.4\textwidth]{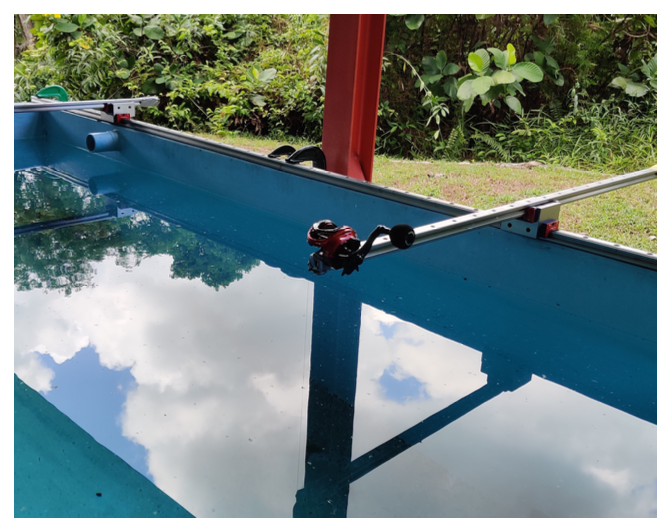}
\end{center}
\caption{Water tank used for the controlled experiment.\label{fig:tank_pic}}
\end{figure}

\subsection{Controlled experiment}\label{sec:controlled-experiment}

With the feasibility established via simulation, we carried out an experimental validation in a water tank using the same setup described in Section~\ref{sec:simulated_feasibility_study}. The equipment setup used in the experiment is shown in Fig.~\ref{fig:tank_flow}. We used a National Instruments Data Acquisition (NI-DAQ) system to transmit a CW signal at $10$~kHz with an amplitude of $1$~V${_\text{pp}}$. A pair of TC4013 acoustic transducers were used as the transmitter and receiver. 500 acoustic measurements were collected at the same locations as the data generated in the feasibility study. Each acoustic transducer was attached to a fishing line, with a reel and sliding block mechanism to control the 3D position of the transducer as shown in Fig.~\ref{fig:tank_pic}. The water tank was located outdoors and experienced \update{a} light breeze on occasion. This led to slight measurement errors due to small-scale oscillations of the source and receiver. The oscillations manifest themselves as fluctuations in the amplitude and phase of the recorded signal. We computed the average envelope over a 40~s period to reduce the impact of oscillations on the measurement. \update{We assign different weights to the 500 measurements to capture the confidence levels of our measurements. The measurements with smaller standard deviations over the 40~s recorded signals weigh more.} The RBNN model allows for errors in direction of arrival to be estimated during training. 

In addition to angular errors, we also expect some errors (\update{a} few cm) in the measurement of the location of the transducers. We design a two-stage training strategy to deal with such location measurement errors. The first stage aims to optimize the trainable parameters $\mathcal{T}$, specified in the designed RBNN model, using measured location data. \update{We train our model using weighted MAE.} We freeze the trained RBNN model at the end of this stage, and focus on estimating measurement errors in the second stage. We feed the corrected locations (measured locations offset by the estimated location errors in all dimensions) into the RBNN model to predict the acoustic field in this stage. A $L_2$-norm penalty term of absolute position errors is added in the loss function to constrain the range of position errors. By minimizing the loss function, the second stage aims to estimate the most appropriate location errors using the RBNN model parameterized by the parameters trained in the first training stage.

\begin{table}[t]
\renewcommand{\arraystretch}{1.25} 
\centering
\caption{\label{tab:tank_err} Performance evaluation of the estimated acoustic field from the controlled experiment.}
\begin{tabular}{lllc}
\toprule \toprule
{Method} & \shortstack{{MATE} \\
 (V$_\text{pp}$)} & \shortstack{ {MATE} \\(dB)} & \shortstack{{Spearman's}\\{correlation coefficient}} \\
\midrule
RBNN & \update{0.003} & \update{0.376} & \update{0.961}\\
GPR  & \update{0.021} & \update{3.764} & \update{-0.021}\\
DNN  &  \update{0.031} & \update{5.316} & \update{-0.207}\\ 
\bottomrule
\end{tabular}
\end{table}

\begin{figure}[t]
\begin{center}
\includegraphics[width=0.45\textwidth]{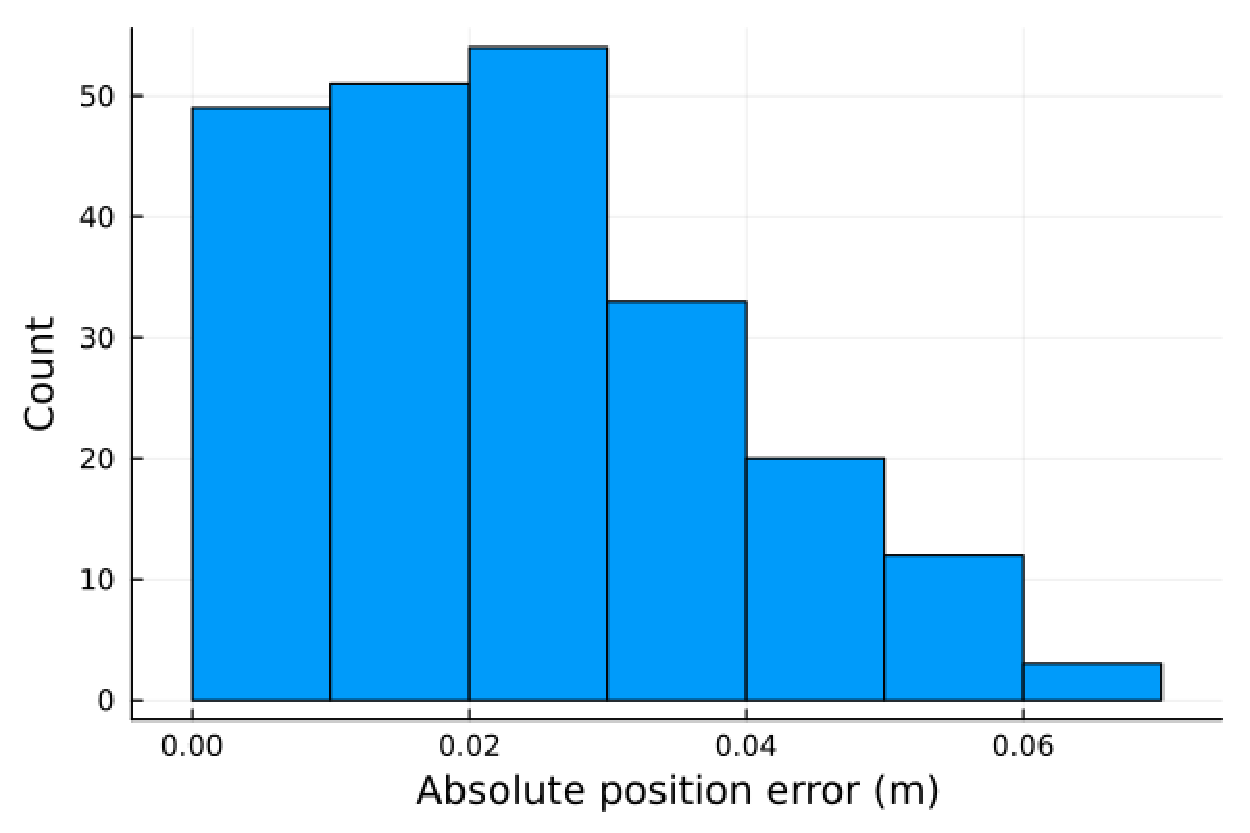}
\end{center}
\caption{\update{Trained absolute position error of the sparse test data using the RBNN model.} \label{fig:test_pos_error}}
\end{figure}

\begin{figure*}
\centering
\begin{subfigure}{\textwidth}
\includegraphics[width=\linewidth]{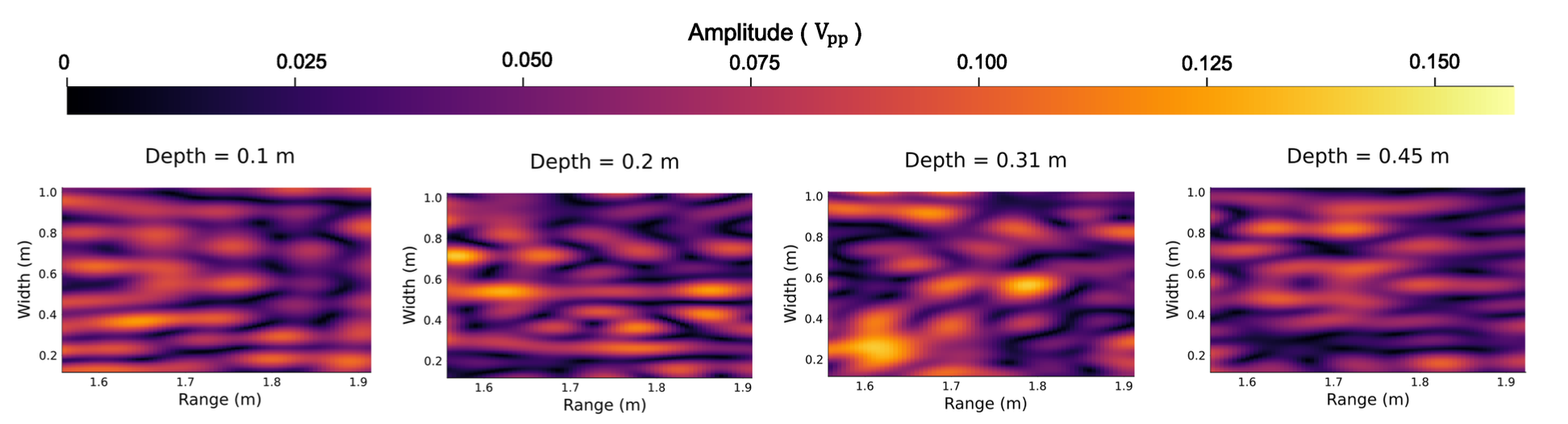}
\caption{RBNN field estimates within AOI}
\vspace{3ex}
\end{subfigure}
\begin{subfigure}{\textwidth}
\includegraphics[width=\linewidth]{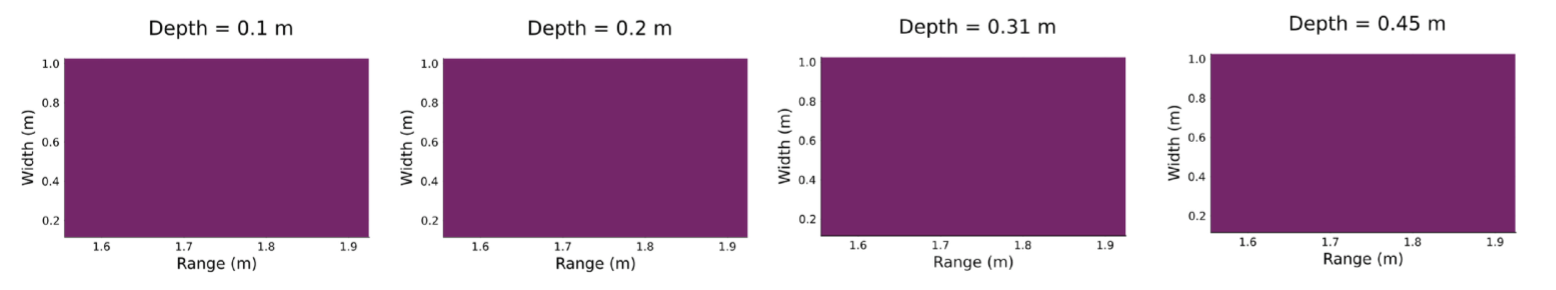}
\caption{GPR field estimates within AOI}
\vspace{3ex}
\end{subfigure}
\begin{subfigure}{\textwidth}
\includegraphics[width=\linewidth]{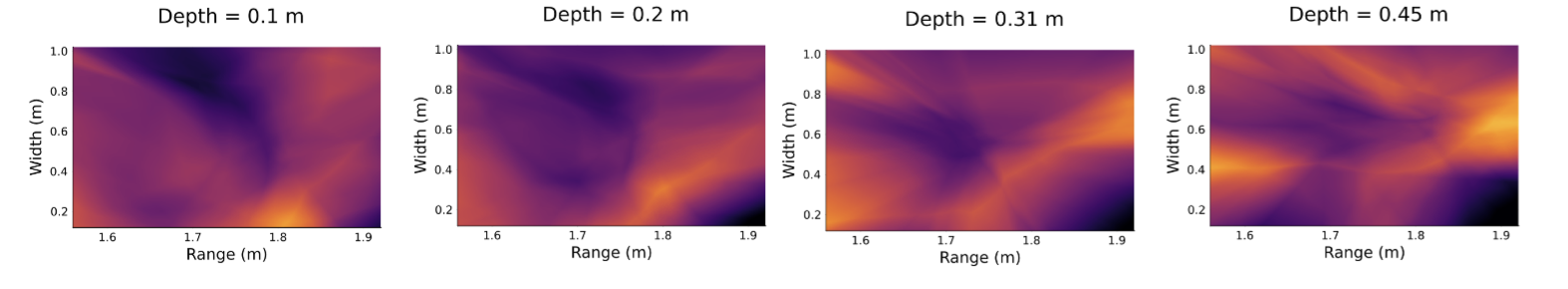}
\caption{DNN field estimates within AOI}
\vspace{3ex}
\end{subfigure}
\caption{\update{Estimated field patterns within AOI using the experimental data by the three models. }}
\label{fig: experiment_local}
\end{figure*}

\begin{figure*}
\begin{subfigure}{\textwidth}
\includegraphics[width=\linewidth]{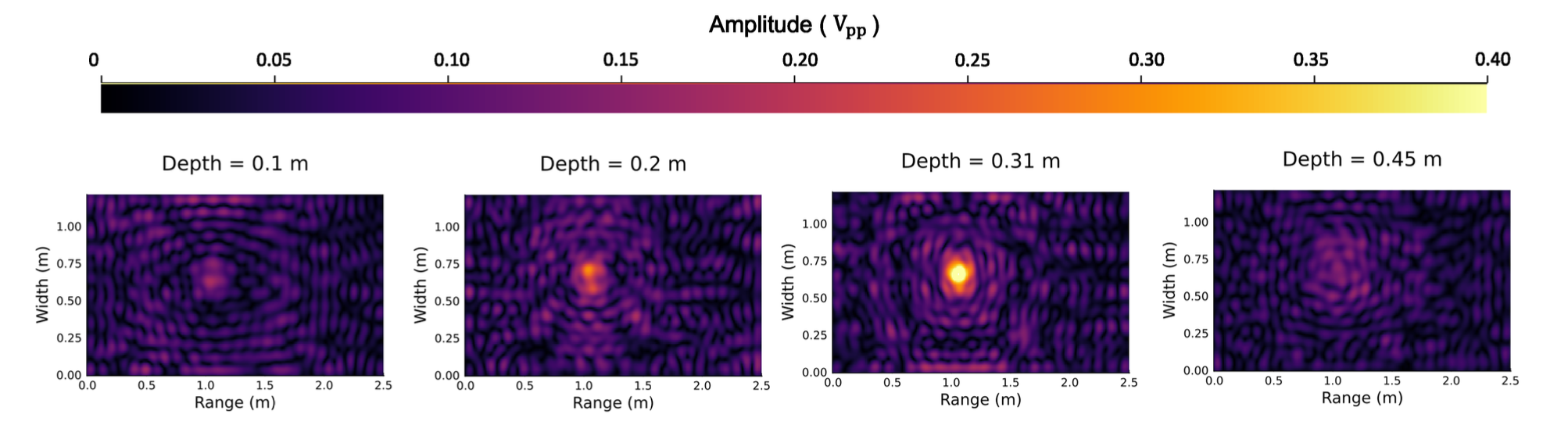}
\caption{RBNN extrapolated field estimates}
\vspace{3ex}
\end{subfigure}
\begin{subfigure}{\textwidth}
\includegraphics[width=\linewidth]{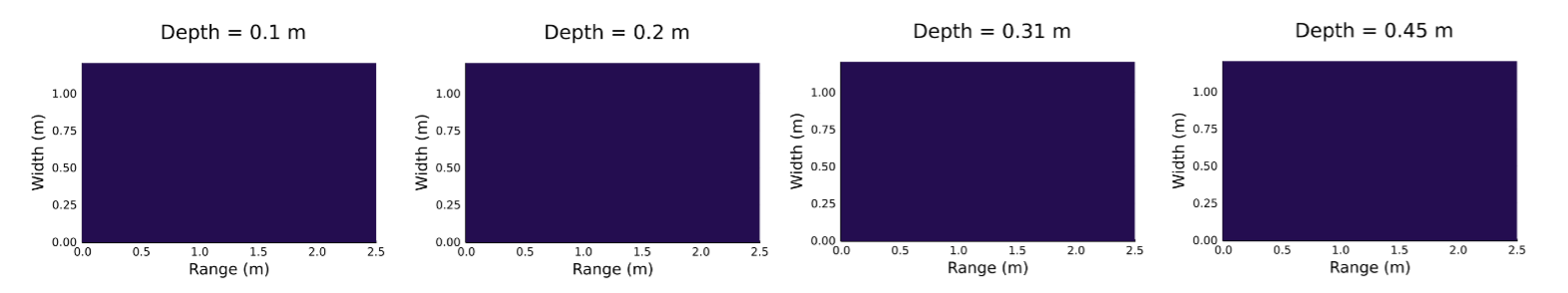}
\caption{GPR extrapolated field estimates}
\vspace{3ex}
\end{subfigure}
\begin{subfigure}{\textwidth}
\includegraphics[width=\linewidth]{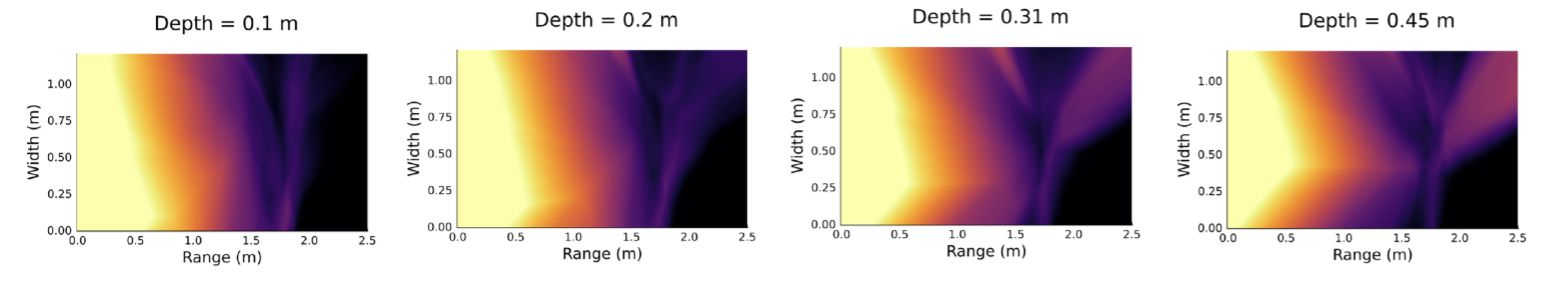}
\caption{DNN extrapolated field estimates}
\vspace{3ex}
\end{subfigure}
\caption{\update{Extrapolated field patterns of the entire tank environment using the experimental data by the three models.}}
\label{fig: experiment_extra}
\end{figure*}

\begin{figure}[t]
\centering
\includegraphics[width=0.45\linewidth]{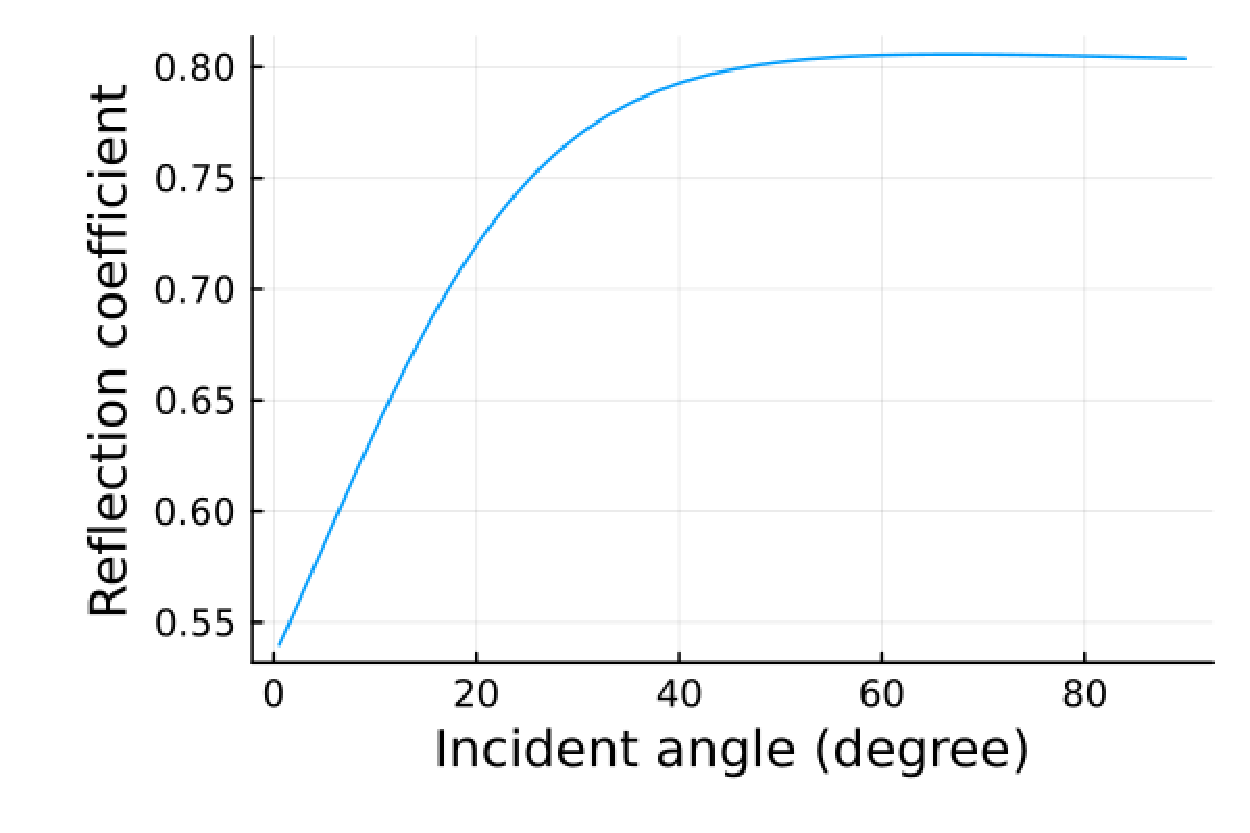}
\caption{\update{Estimated reflection coefficient based on the trained RCNN layer of the composite RBNN.}}
\label{fig:learnt_rcnn} 
\end{figure}

\begin{figure*}[t]
\begin{center}
\includegraphics[width=0.9\textwidth]{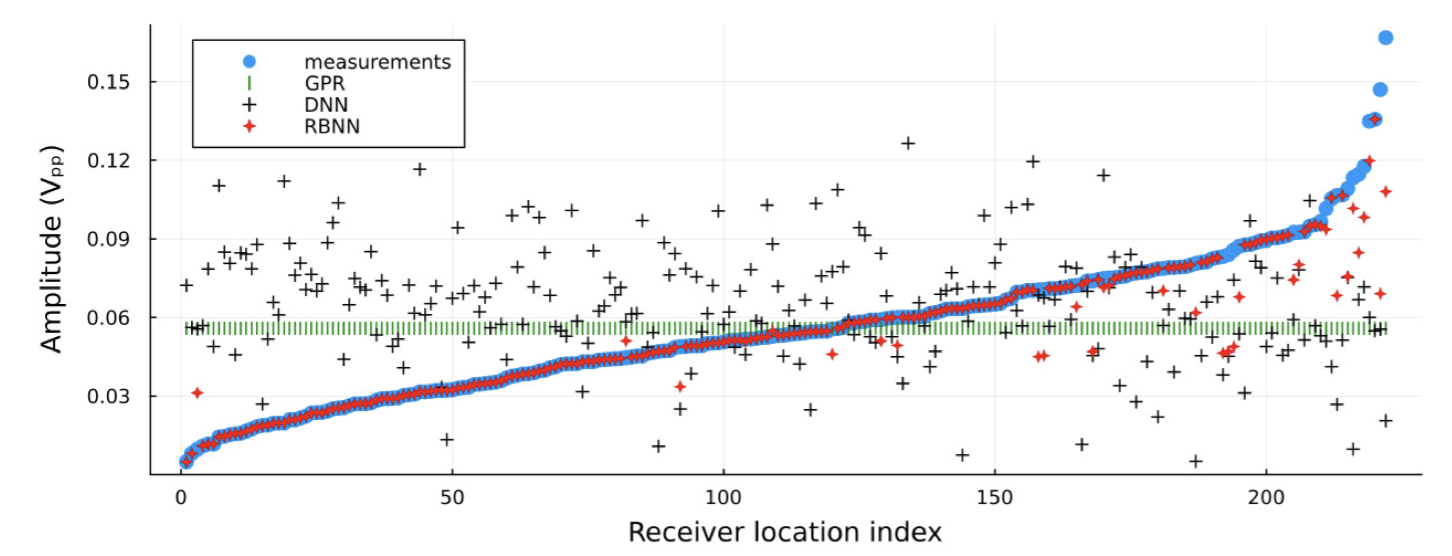}
\end{center}
\caption{\update{Comparison between test data and model predictions}}\label{fig:prediction_vs_mea}
\end{figure*}

To benchmark the RBNN performance, we use a GPR \update{model} and \update{a} DNN \update{model}\footnote{We found that the DNN performed better with experimental data if we replaced the ReLU activation function with a hyperbolic tangent~(tanh) activation function, and therefore we present results for the tanh-activated DNN in this section.} as in the feasibility study. For each of the three methods, Figs.~\ref{fig: experiment_local} and~\ref{fig: experiment_extra} show the estimated fields within the AOI and the extrapolated fields in the entire tank respectively. The field pattern extrapolated by the RBNN model looks reasonable in the sense that the region with the strongest pressure amplitude is consistent with the source location. The GPR and DNN fail to reconstruct any discernable field pattern in the tank. In line with this, the MATE of the RBNN model is significantly lower than that of the GPR and DNN models, as shown in Table~\ref{tab:tank_err}. The absolute trained position errors for the 222 sparse test data points are shown in Fig.~\ref{fig:test_pos_error}. Most errors are below 4~cm, as we would expect from our measurement procedure.

Fig.~\ref{fig:learnt_rcnn} shows the \update{learned} reflection coefficient for the water tank walls. While we do not have ground truth to validate the reflection coefficient curve, the \update{learned} model works well to estimate the acoustic field in the tank. We observe this in the good agreement between RBNN prediction and measured data in Fig.~\ref{fig:prediction_vs_mea}, and also as a Spearman's correlation coefficient of \update{0.961} between the prediction and data in Table~\ref{tab:tank_err}. On the other hand, the GPR and DNN simply learn to predict average values regardless of the measurement location. The results obtained from the controlled experiment thus validate the efficacy of our proposed method to model acoustic propagation in unknown or partially known environments.

\section{Discussion}\label{sec:discussion}

\update{If perfect environmental knowledge is available, one can use physics-based models for field estimation. In the absence of environmental knowledge and when field interpolation is of interest, data-driven methods can be employed to interpolate fields at unvisited locations. Our proposed method is particularly advantageous in practical scenarios where only partial environmental knowledge is available and extrapolation is required. It enables field interpolation and extrapolation using a limited number of acoustic observations by incorporating the available domain knowledge.}

The RBNN framework offers a high-frequency physics-based acoustic propagation modeling approach that can incorporate known environmental information and be trained with observed data, making it suitable for solving \update{acoustic modeling} problems with limited data. The approach that we took to derive the RBNN from a high-frequency approximation to the solution of a wave equation can also be applied to other approximations. For example, applying the same approach to a normal-mode approximation \update{yields} normal-mode neural networks that may be used for low-frequency acoustic propagation problems~\update{\cite{li2022physics}}. The RBNN framework \update{could} incorporate environmental complexities such as range-dependent bathymetry, non-isovelocity sound speed profiles, and various geo-acoustic models. When the physics is fully or partially known, explicit expressions can be included in the computational graph, with potentially some unknown parameters. On the other hand, when physics is unknown, \update{NNs} can be used as components of the computational graph to model arbitrary functions. The resulting computational graph can be automatically differentiated with respect to the model parameters, thus making it suitable for training with standard \update{gradient-descent-based NN} training algorithms.

\update{The specific components that constitute RBNN trainable parameters are determined by the problem formulation. There are several factors that can affect the model training time, including the number of training epochs allowed, the size of training data and the size of trainable parameters. In general, on typical personal computers today, it takes at most a few minutes to train the three models for studies evaluated in this paper. For example, on a MacBook Air with an M2 chip and 16~GB RAM, it takes 58.8~s, 2.23~ms and 31.9~s\footnote{\update{We measure the minimum elapsed time during the benchmark for the three models.}} to train the RBNN model, GPR model and DNN model respectively for the near field estimation problem in Section~\ref{sec:sph_field_est}.}

\update{Our method is robust to noise, as the loss function minimizes noise regardless of its level. Our approach can be applied at various propagation ranges as we make no implicit assumptions about it. Typically field patterns become less complicated with range, and so become easier to model. The approach can be applied across a vast range of frequencies. However, we require more training data for higher frequencies, as the field structure becomes more intricate at smaller scales as frequencies increase.}

\update{In our current work, we assumed quasi-static environments. This assumption could lead to serious performance degradation in environments that undergo significant changes over the period during which acoustic observations are collected, particularly at high frequencies. One of the possible ways to address environmental dynamics is to incorporate the knowledge of environmental changes into the modeling. This would allow us to generate a time-varying physics-based data-aided propagation model. For example, if an environmental parameter undergoes a first-order linear change over time, one could model RBNN parameters as linear functions of time.}

\section{\update{Conclusions}}\label{sec:conclusion}

\update{Modeling acoustic propagation in underwater environments is particularly challenging in partially unknown environments. Physics-based acoustic propagation models have limited practical uses as they require accurate prior environmental knowledge to estimate acoustic fields in a given area.} While classical \update{ML} techniques such as GPR and DNN can use data to approximate an unknown acoustic propagation model, they lack the ability to incorporate scientific domain knowledge (in our case, the acoustic wave equation) and environmental knowledge (e.g. known channel geometry). Without incorporating the domain or environmental knowledge, these techniques tend to be data-hungry during training, and extrapolate poorly. They can also make predictions that are physically unrealistic.

\update{To address the limitations in physics-based models and data-driven modeling techniques, we proposed a modeling recipe that embeds scientific domain knowledge into data-driven ML to leverage their complementary strengths.} We showed that \update{our} RBNN method can learn from very little data, extrapolate well beyond the region where training data \update{are} available, and always make predictions that are consistent with physics. We demonstrated a few applications of the RBNN framework, highlighting the flexibility it provides in modeling acoustic propagation scenarios with varying degrees of environmental complexity and knowledge. We believe that the framework can be applied to solve a much wider variety of acoustic propagation problems. 

\bibliographystyle{ieeetr} 
\bibliography{biblio}

\begin{thebibliography}{10}

\bibitem{james2008method}
K.~R. James and D.~R. Dowling, ``A method for approximating
  acoustic-field-amplitude uncertainty caused by environmental uncertainties,''
  {\em The Journal of the Acoustical Society of America}, vol.~124, no.~3,
  pp.~1465--1476, 2008.

\bibitem{gul2017underwater}
S.~Gul, S.~S.~H. Zaidi, R.~Khan, and A.~B. Wala, ``Underwater acoustic channel
  modeling using {BELLHOP} ray tracing method,'' in {\em 14th International
  Bhurban Conference on Applied Sciences and Technology (IBCAST)},
  pp.~665--670, IEEE, 2017.

\bibitem{llor2012underwater}
J.~Llor and M.~P. Malumbres, ``Underwater wireless sensor networks: {H}ow do
  acoustic propagation models impact the performance of higher-level
  protocols?,'' {\em Sensors}, vol.~12, no.~2, pp.~1312--1335, 2012.

\bibitem{Chapman2008}
N.~R. Chapman, ``Inverse problems in underwater acoustics,'' in {\em Handbook
  of Signal Processing in Acoustics}, pp.~1723--1735, Springer New York, 2008.

\bibitem{chapman2003benchmarking}
N.~R. Chapman, S.~Chin-Bing, D.~King, and R.~B. Evans, ``Benchmarking
  geoacoustic inversion methods for range-dependent waveguides,'' {\em IEEE
  journal of oceanic engineering}, vol.~28, no.~3, pp.~320--330, 2003.

\bibitem{dosso1993estimation}
S.~Dosso, M.~Yeremy, J.~Ozard, and N.~Chapman, ``Estimation of ocean-bottom
  properties by matched-field inversion of acoustic field data,'' {\em IEEE
  Journal of Oceanic Engineering}, vol.~18, no.~3, pp.~232--239, 1993.

\bibitem{bonnel2010estimation}
J.~Bonnel, B.~Nicolas, J.~I. Mars, and S.~C. Walker, ``Estimation of modal
  group velocities with a single receiver for geoacoustic inversion in shallow
  water,'' {\em The Journal of the Acoustical Society of America}, vol.~128,
  no.~2, pp.~719--727, 2010.

\bibitem{collins1994inverse}
M.~Collins and W.~Kuperman, ``Inverse problems in ocean acoustics,'' {\em
  Inverse Problems}, vol.~10, no.~5, p.~1023, 1994.

\bibitem{tolstoy2000applications}
A.~Tolstoy, ``Applications of matched-field processing to inverse problems in
  underwater acoustics,'' {\em Inverse Problems}, vol.~16, no.~6, p.~1655,
  2000.

\bibitem{baggeroer1988matched}
A.~B. Baggeroer, W.~Kuperman, and H.~Schmidt, ``Matched field processing:
  Source localization in correlated noise as an optimum parameter estimation
  problem,'' {\em The Journal of the Acoustical Society of America}, vol.~83,
  no.~2, pp.~571--587, 1988.

\bibitem{baggeroer1993overview}
A.~B. Baggeroer, W.~A. Kuperman, and P.~N. Mikhalevsky, ``An overview of
  matched field methods in ocean acoustics,'' {\em IEEE Journal of Oceanic
  Engineering}, vol.~18, no.~4, pp.~401--424, 1993.

\bibitem{Jensen2011propagation}
F.~B. Jensen, W.~A. Kuperman, M.~B. Porter, and H.~Schmidt, ``Wave propagation
  theory,'' in {\em Computational Ocean Acoustics}, pp.~65--153, Springer New
  York, 2011.

\bibitem{10.3389/fmars.2021.751327}
T.~C.~A. Oliveira, Y.-T. Lin, and M.~B. Porter, ``Underwater sound propagation
  modeling in a complex shallow water environment,'' {\em Frontiers in Marine
  Science}, vol.~8, p.~1464, 2021.

\bibitem{Jensen2011}
F.~B. Jensen, W.~A. Kuperman, M.~B. Porter, and H.~Schmidt, ``Ray methods,'' in
  {\em Computational Ocean Acoustics}, pp.~155--232, Springer New York, 2011.

\bibitem{Jensen2011nm}
F.~B. Jensen, W.~A. Kuperman, M.~B. Porter, and H.~Schmidt, ``Normal modes,''
  in {\em Computational Ocean Acoustics}, pp.~337--455, Springer New York,
  2011.

\bibitem{Jensen2011pb}
F.~B. Jensen, W.~A. Kuperman, M.~B. Porter, and H.~Schmidt, ``Parabolic
  equations,'' in {\em Computational Ocean Acoustics}, pp.~457--529, Springer
  New York, 2011.

\bibitem{Jensen2011wni}
F.~B. Jensen, W.~A. Kuperman, M.~B. Porter, and H.~Schmidt, ``Wavenumber
  integration techniques,'' in {\em Computational Ocean Acoustics},
  pp.~233--335, Springer New York, 2011.

\bibitem{williamson1995critical}
P.~R. Williamson and R.~G. Pratt, ``A critical review of acoustic wave modeling
  procedures in 2.5 dimensions,'' {\em Geophysics}, vol.~60, no.~2,
  pp.~591--595, 1995.

\bibitem{reeder2010experimental}
D.~B. Reeder, L.~Y. Chiu, and C.-F. Chen, ``Experimental evidence of horizontal
  refraction by nonlinear internal waves of elevation in shallow water in the
  south china sea: 3{D} versus {N}x2{D} acoustic propagation modeling,'' {\em
  Journal of Computational Acoustics}, vol.~18, no.~03, pp.~267--278, 2010.

\bibitem{wang2014review}
L.~Wang, K.~Heaney, T.~Pangerc, P.~Theobald, S.~Robinson, and M.~Ainslie,
  ``Review of underwater acoustic propagation models,'' tech. rep., National
  Physical Laboratory, 2014.

\bibitem{DiNapoli1979}
F.~R. DiNapoli and R.~L. Deavenport, ``Numerical models of underwater acoustic
  propagation,'' in {\em Ocean Acoustics}, pp.~79--157, Springer Berlin
  Heidelberg, 1979.

\bibitem{5422478}
P.~C. Etter, ``Review of ocean-acoustic models,'' in {\em OCEANS 2009},
  pp.~1--6, 2009.

\bibitem{goodfellow2016machine}
I.~Goodfellow, Y.~Bengio, and A.~Courville, ``Machine learning basics,'' {\em
  Deep learning}, vol.~1, no.~7, pp.~96--160, 2016.

\bibitem{jordan2015machine}
M.~I. Jordan and T.~M. Mitchell, ``Machine learning: Trends, perspectives, and
  prospects,'' {\em Science}, vol.~349, no.~6245, pp.~255--260, 2015.

\bibitem{caviedes2021gaussian}
D.~Caviedes-Nozal, N.~A. Riis, F.~M. Heuchel, J.~Brunskog, P.~Gerstoft, and
  E.~Fernandez-Grande, ``Gaussian processes for sound field reconstruction,''
  {\em The Journal of the Acoustical Society of America}, vol.~149, no.~2,
  pp.~1107--1119, 2021.

\bibitem{kohlsche2019gaussian}
T.~Kohlsche, S.~Lippert, and O.~von Estorff, ``Gaussian process based surrogate
  modelling of acoustic systems,'' {\em Proceedings in Applied Mathematics and
  Mechanics}, vol.~19, no.~1, 2019.

\bibitem{lee2022predicting}
B.~M. Lee, J.~R. Johnson, and D.~R. Dowling, ``Predicting acoustic transmission
  loss uncertainty in ocean environments with neural networks,'' {\em Journal
  of Marine Science and Engineering}, vol.~10, no.~10, p.~1548, 2022.

\bibitem{mallik2022predicting}
W.~Mallik, R.~K. Jaiman, and J.~Jelovica, ``Predicting transmission loss in
  underwater acoustics using convolutional recurrent autoencoder network,''
  {\em The Journal of the Acoustical Society of America}, vol.~152, no.~3,
  pp.~1627--1638, 2022.

\bibitem{bernardo1998regression}
J.~Bernardo, J.~Berger, A.~Dawid, A.~Smith, {\em et~al.}, ``Regression and
  classification using {G}aussian process priors,'' {\em Bayesian statistics},
  vol.~6, p.~475, 1998.

\bibitem{Rasmussen2004}
C.~E. Rasmussen, ``Gaussian processes in machine learning,'' in {\em Advanced
  Lectures on Machine Learning: ML Summer Schools}, pp.~63--71, 2004.

\bibitem{seeger2004gaussian}
M.~Seeger, ``Gaussian processes for machine learning,'' {\em International
  journal of neural systems}, vol.~14, no.~02, pp.~69--106, 2004.

\bibitem{murphy2022probabilistic}
K.~P. Murphy, ``Gaussian processes,'' in {\em Probabilistic machine learning:
  an introduction}, pp.~515--542, MIT press, 2022.

\bibitem{liu2020gaussian}
H.~Liu, Y.-S. Ong, X.~Shen, and J.~Cai, ``When {G}aussian process meets big
  data: A review of scalable {GP}s,'' {\em IEEE transactions on neural networks
  and learning systems}, vol.~31, no.~11, pp.~4405--4423, 2020.

\bibitem{gurney2018introduction}
K.~Gurney, ``Neural networks---an overview,'' in {\em An introduction to neural
  networks}, pp.~1--4, CRC press, 2018.

\bibitem{hornik1989multilayer}
K.~Hornik, M.~Stinchcombe, and H.~White, ``Multilayer feedforward networks are
  universal approximators,'' {\em Neural networks}, vol.~2, no.~5,
  pp.~359--366, 1989.

\bibitem{2017}
A.~Karpatne, G.~Atluri, J.~H. Faghmous, M.~Steinbach, A.~Banerjee, A.~Ganguly,
  S.~Shekhar, N.~Samatova, and V.~Kumar, ``Theory-guided data science: A new
  paradigm for scientific discovery from data,'' {\em IEEE Transactions on
  Knowledge and Data Engineering}, vol.~29, pp.~2318--2331, Oct 2017.

\bibitem{osti_1478744}
N.~Baker, F.~Alexander, T.~Bremer, A.~Hagberg, Y.~Kevrekidis, H.~Najm,
  M.~Parashar, A.~Patra, J.~Sethian, S.~Wild, {\em et~al.}, ``Workshop report
  on basic research needs for scientific machine learning: Core technologies
  for artificial intelligence,'' tech. rep., USDOE Office of Science (SC),
  Washington, DC (United States), 2019.

\bibitem{raissi2018hidden}
M.~Raissi and G.~E. Karniadakis, ``Hidden physics models: Machine learning of
  nonlinear partial differential equations,'' {\em Journal of Computational
  Physics}, vol.~357, pp.~125--141, 2018.

\bibitem{swiler2020survey}
L.~P. Swiler, M.~Gulian, A.~L. Frankel, C.~Safta, and J.~D. Jakeman, ``A survey
  of constrained {G}aussian process regression: Approaches and implementation
  challenges,'' {\em Journal of Machine Learning for Modeling and Computing},
  vol.~1, no.~2, 2020.

\bibitem{willard2021integrating}
J.~Willard, X.~Jia, S.~Xu, M.~Steinbach, and V.~Kumar, ``Integrating scientific
  knowledge with machine learning for engineering and environmental systems,''
  2021.

\bibitem{DBLP:journals/corr/abs-2001-04385}
C.~Rackauckas, Y.~Ma, J.~Martensen, C.~Warner, K.~Zubov, R.~Supekar,
  D.~Skinner, and A.~J. Ramadhan, ``Universal differential equations for
  scientific machine learning,'' {\em CoRR}, vol.~abs/2001.04385, 2020.

\bibitem{read2019process}
J.~S. Read, X.~Jia, J.~Willard, A.~P. Appling, J.~A. Zwart, S.~K. Oliver,
  A.~Karpatne, G.~J. Hansen, P.~C. Hanson, W.~Watkins, {\em et~al.},
  ``Process-guided deep learning predictions of lake water temperature,'' {\em
  Water Resources Research}, vol.~55, no.~11, pp.~9173--9190, 2019.

\bibitem{sun2019theory}
J.~Sun, Z.~Niu, K.~A. Innanen, J.~Li, and D.~O. Trad, ``A theory-guided deep
  learning formulation of seismic waveform inversion,'' in {\em SEG Technical
  Program Expanded Abstracts 2019}, pp.~2343--2347, Society of Exploration
  Geophysicists, 2019.

\bibitem{raissi2019physics}
M.~Raissi, P.~Perdikaris, and G.~E. Karniadakis, ``Physics-informed neural
  networks: A deep learning framework for solving forward and inverse problems
  involving nonlinear partial differential equations,'' {\em Journal of
  Computational Physics}, vol.~378, pp.~686--707, 2019.

\bibitem{borrel2021physics}
N.~Borrel-Jensen, A.~P. Engsig-Karup, and C.-H. Jeong, ``Physics-informed
  neural networks for one-dimensional sound field predictions with
  parameterized sources and impedance boundaries,'' {\em {JASA} Express
  Letters}, vol.~1, no.~12, p.~122402, 2021.

\bibitem{moseley2020solving}
B.~Moseley, A.~Markham, and T.~Nissen-Meyer, ``Solving the wave equation with
  physics-informed deep learning,'' {\em arXiv preprint arXiv:2006.11894},
  2020.

\bibitem{rasht2021physics}
M.~Rasht-Behesht, C.~Huber, K.~Shukla, and G.~E. Karniadakis,
  ``Physics-informed neural networks ({PINN}s) for wave propagation and full
  waveform inversions,'' {\em arXiv preprint arXiv:2108.12035}, 2021.

\bibitem{de2021assessing}
T.~de~Wolff, H.~Carrillo, L.~Mart{\'\i}, and N.~Sanchez-Pi, ``Assessing physics
  informed neural networks in ocean modelling and climate change
  applications,'' in {\em AI: Modeling Oceans and Climate Change Workshop at
  ICLR 2021}, 2021.

\bibitem{de2021towards}
T.~de~Wolff, H.~Carrillo, L.~Mart{\'\i}, and N.~Sanchez-Pi, ``Towards optimally
  weighted physics-informed neural networks in ocean modelling,'' {\em arXiv
  preprint arXiv:2106.08747}, 2021.

\bibitem{kexin2021ocean}
K.~Li and M.~Chitre, ``Ocean acoustic propagation modeling using scientific
  machine learning,'' in {\em OCEANS 2021: San Diego--Porto}, pp.~1--5, IEEE,
  2021.

\bibitem{li2022physics}
K.~Li and M.~A. Chitre, ``Physics-aided data-driven modal ocean acoustic
  propagation modeling,'' in {\em 24th International Congress on Acoustics
  (ICA)}, pp.~1--9, 2022.

\bibitem{hovem2013ray}
J.~M. Hovem, ``Ray trace modeling of underwater sound propagation,'' in {\em
  Modeling and measurement methods for acoustic waves and for acoustic
  microdevices}, IntechOpen, 2013.

\bibitem{hecht1992theory}
R.~Hecht-Nielsen, ``Theory of the backpropagation neural network,'' in {\em
  Neural networks for perception}, pp.~65--93, Elsevier, 1992.

\bibitem{werbos1974beyond}
P.~Werbos, ``Beyond regression: New tools for prediction and analysis in the
  behavioral sciences,'' {\em PhD thesis, Committee on Applied Mathematics,
  Harvard University, Cambridge, MA}, 1974.

\bibitem{baydin2018automatic}
A.~G. Baydin, B.~A. Pearlmutter, A.~A. Radul, and J.~M. Siskind, ``Automatic
  differentiation in machine learning: {A} survey,'' {\em Journal of machine
  learning research}, vol.~18, 2018.

\bibitem{Kingma2015AdamAM}
D.~P. Kingma and J.~Ba, ``Adam: A method for stochastic optimization,'' {\em
  CoRR}, vol.~abs/1412.6980, 2015.

\bibitem{ketkar2017stochastic}
N.~Ketkar and N.~Ketkar, ``Stochastic gradient descent,'' {\em Deep learning
  with Python: A hands-on introduction}, pp.~113--132, 2017.

\bibitem{berahas2016multi}
A.~S. Berahas, J.~Nocedal, and M.~Tak{\'a}c, ``A multi-batch {L-BFGS} method
  for machine learning,'' {\em Advances in Neural Information Processing
  Systems}, vol.~29, 2016.

\bibitem{li2014efficient}
M.~Li, T.~Zhang, Y.~Chen, and A.~J. Smola, ``Efficient mini-batch training for
  stochastic optimization,'' in {\em Proceedings of the 20th ACM SIGKDD
  international conference on Knowledge discovery and data mining},
  pp.~661--670, 2014.

\bibitem{Jensen2011fun}
F.~B. Jensen, W.~A. Kuperman, M.~B. Porter, and H.~Schmidt, ``Fundamentals of
  ocean acoustics,'' in {\em Computational Ocean Acoustics}, pp.~1--64,
  Springer New York, 2011.

\bibitem{Kistovich2020}
A.~Kistovich, K.~Pokazeev, and T.~Chaplina, ``General properties and character
  types of sound waves,'' in {\em Ocean Acoustics}, pp.~23--41, Springer
  International Publishing, 2020.

\bibitem{morse1986theoretical}
P.~M. Morse and K.~U. Ingard, ``The radiation of sound,'' in {\em Theoretical
  acoustics}, pp.~306--394, Princeton university press, 1986.

\bibitem{fisher1977sound}
F.~Fisher and V.~Simmons, ``Sound absorption in sea water,'' {\em The Journal
  of the Acoustical Society of America}, vol.~62, no.~3, pp.~558--564, 1977.

\bibitem{Brekhovskikh2003}
L.~Brekhovskikh and Y.~P. Lysanov, ``Reflection of sound from the surface and
  bottom of the ocean: Plane waves,'' in {\em Fundamentals of Ocean Acoustics},
  pp.~61--79, Springer New York, 2003.

\bibitem{allen1979image}
J.~B. Allen and D.~A. Berkley, ``Image method for efficiently simulating
  small-room acoustics,'' {\em The Journal of the Acoustical Society of
  America}, vol.~65, no.~4, pp.~943--950, 1979.

\bibitem{jackson1994apl}
APL-UW, ``Bottom,'' in {\em High-Frequency Ocean Environmental Acoustic Models
  Handbook}, pp.~122--175, Washington Univ Seattle Applied Physics Lab, 1994.

\bibitem{prechelt2012early}
L.~Prechelt, ``Early stopping--but when?,'' in {\em Neural networks: tricks of
  the trade: second edition}, pp.~53--67, Springer, 2012.

\bibitem{porter2011bellhop}
M.~B. Porter, ``The {BELLHOP} manual and user's guide: Preliminary draft,''
  {\em Heat, Light, and Sound Research}, vol.~260, 2011.

\bibitem{mandar_pekeris}
M.~Chitre, ``Differentiable ocean acoustic propagation modeling,'' in {\em
  IEEE/MTS OCEANS Limerick}, 2023.

\end{thebibliography}

\end{document}